\documentclass[astrosymb, twocolumn]{aastex631}

\received{Sep 2, 2022}
\revised{Feb 3, 2023}
\accepted{Feb 17, 2023}

\submitjournal{ApJ}

\shorttitle{Makani Ionization}
\shortauthors{Rupke et al.}

\graphicspath{{./}{figures/}}

\begin{document}

\title{The Ionization and Dynamics of the Makani Galactic Wind}

\correspondingauthor{David Rupke}
\email{drupke@gmail.com}

\author[0000-0002-1608-7564]{David S. N. Rupke}
\affiliation{Department of Physics, Rhodes College, 2000 North Parkway, Memphis, TN 38112, USA}

\author{Alison L. Coil}
\affil{Center for Astrophysics and Space Sciences, University of California, San Diego, La Jolla, CA 92093, USA}

\author{Serena Perrotta}
\affil{Center for Astrophysics and Space Sciences, University of California, San Diego, La Jolla, CA 92093, USA}

\author{Julie D. Davis}
\affil{Department of Astronomy, University of Wisconsin-Madison, Madison, WI 53706, USA}

\author{Aleksandar M. Diamond-Stanic}
\affil{Department of Physics and Astronomy, Bates College, Lewiston, ME, 04240, USA}

\author{James E. Geach}
\affil{Centre for Astrophysics Research, University of Hertfordshire, Hatfield, Hertfordshire AL10 9AB, UK}

\author{Ryan C. Hickox}
\affil{Department of Physics and Astronomy, Dartmouth College, Hanover, NH 03755, USA}

\author{John Moustakas}
\affil{Department of Physics and Astronomy, Siena College, Loudonville, NY 12211, USA}

\author{Grayson C. Petter}
\affil{Department of Physics and Astronomy, Dartmouth College, Hanover, NH 03755, USA}

\author{Gregory H. Rudnick}
\affil{Department of Physics and Astronomy, University of Kansas, Lawrence, KS 66045, USA}

\author{Paul H. Sell}
\affil{Department of Astronomy, University of Florida, Gainesville, FL, 32611 USA}

\author{Christy A. Tremonti}
\affil{Department of Astronomy, University of Wisconsin-Madison, Madison, WI 53706, USA}

\author{Kelly E. Whalen}
\affil{Department of Physics and Astronomy, Dartmouth College, Hanover, NH 03755, USA}

\begin{abstract}

The Makani galaxy hosts the poster child of a galactic wind on scales of the circumgalactic medium. It consists of a two-episode wind in which the slow, outer wind originated 400~Myr ago (Episode I; $R_\mathrm{I}=20-50$~kpc) and the fast, inner wind is 7~Myr old (Episode II; $R_\mathrm{II}=0-20$~kpc). While this wind contains ionized, neutral, and molecular gas, the physical state and mass of the most extended phase---the warm, ionized gas---is unknown. Here we present Keck optical spectra of the Makani outflow. These allow us to detect hydrogen lines out to $r=30-40$~kpc and thus constrain the mass, momentum, and energy in the wind. Many collisionally-excited lines are detected throughout the wind, and their line ratios are consistent with 200--400~km~s$^{-1}$ shocks that power the ionized gas, with $v_\mathrm{shock} = \sigma_\mathrm{wind}$. Combining shock models, density-sensitive line ratios, and mass and velocity measurements, we estimate that the ionized mass and outflow rate in the Episode II wind could be as high as that of the molecular gas: $M_\mathrm{II}^\mathrm{HII}\sim M_\mathrm{II}^\mathrm{H_2}=(1-2)\times10^9~\mathrm{M}_\odot$ and $dM/dt_\mathrm{II}^\mathrm{HII}\sim dM/dt_\mathrm{II}^{\mathrm{H}_2}=170-250~\mathrm{M}_\odot$~yr$^{-1}$. The outer wind has slowed, so that $dM/dt_\mathrm{I}^\mathrm{HII} \sim 10~\mathrm{M}_\odot$~yr$^{-1}$, but it contains more ionized gas: $M_\mathrm{I}^\mathrm{HII}=5\times10^9$~M$_\odot$. The momentum and energy in the recent Episode~II wind imply a momentum-driven flow ($p$ ``boost" $\sim$7) driven by the hot ejecta and radiation pressure from the Eddington-limited, compact starburst. Much of the energy and momentum in the older Episode I wind may reside in a hotter phase, or lie further into the CGM.

\end{abstract}


\section{Introduction} \label{sec:intro}

The circumgalactic medium (CGM) of galaxies contains at least 80\%\ of the baryonic mass within their dark matter halos \citep{2012ApJ...759...23S,2014ApJ...792....8W,2017ARA&A..55..389T}. The CGM is also home to $>$50\%\ of the metals within a halo \citep{2014ApJ...786...54P,2014ApJ...796..136B,2016MNRAS.460.2157O}. The origin of this gas is likely diverse, but an important source is metal-enriched gas ejected from galaxies by galactic winds.

Catching galactic winds in the act of depositing gas in the circumgalactic medium has proven challenging, as the observed sizes of these winds are tpyically comparable to the scales of the galaxies themselves (of order $\la$10~kpc). We reported in \citet{2019Natur.574..643R} a 100~kpc nebula surrounding the Makani galaxy at $z=0.459$. The galaxy Makani (SDSS J211824.06+001729.4) is a massive ($M_\odot = 10^{11.1}$) star-forming galaxy with $r_e = 2.5$~kpc \citep{2014MNRAS.441.3417S}. The nebula, observed in [\ion{O}{2}]$~3726,~3729$~\AA, is consistent with two galactic wind episodes over the past $\sim$400~Myr, based on analysis of its morphology, kinematics, and stellar populations. Episode I was powered by a star formation episode 400~Myr in the past, and includes most of the outer 20--50~kpc of the wind. This wind has slow projected speeds $\sim$100~km~s$^{-1}$ and linewidths $\sigma = 200$~km~$^{-1}$, with a shape characteristic of a giant, bipolar outflow. The star formation timescale, projected ballistic flow speed, and radius of this flow are consistent: $\langle v \rangle_\mathrm{I} \sim R_\mathrm{I}/t_\mathrm{*,I} = 50~\mathrm{kpc}/400~\mathrm{Myr} = 120$~km~s$^{-1}$. Episode II was powered by star formation a mere 7~Myr ago, and consists of a fast wind with maximum speeds exceeding 2000~km~s$^{-1}$, similar to the extended, high-velocity flow seen in other compact starbursts \citep{2014Natur.516...68G}. Most of the Episode II wind is within 20~kpc of the host galaxy, though there is a faint southern extension to 40~kpc. As with Episode I, the approximate timescale, speed, and size line up: $\langle v \rangle_\mathrm{II} \sim R_\mathrm{II}/t_\mathrm{*,II} = 10~\mathrm{kpc}/7~\mathrm{Myr} = 1400$~km~s$^{-1}$. The overall size of the Episode I$+$II nebula relative to the parent galaxy ($r_\mathrm{wind}/r_\mathrm{e,*}\ga20$) is direct evidence that the wind has moved into the galaxy's CGM.

The KCWI observations covered only blue wavelengths and emission lines from [\ion{O}{2}], \ion{Mg}{2}~2796,~2803~\AA, and [\ion{Ne}{5}]~3345,~3426~\AA. They thus leave unanswered two important questions regarding the ionized gas in the Makani wind. First, the ionized mass in the wind cannot be determined without recombination-line measurements. Second, the ionization state of the wind is unknown. The former is critical for constraining the impact of the wind on its host and surroundings; the second is important for understanding its interaction with the CGM. An understanding of how the nebula is powered may also inform how it is driven.

The spatially-resolved molecular component of the Episode II wind in Makani is outflowing with similar velocities to the ionized gas at a rate of 245~M$_\odot$~yr$^{-1}$ \citep{2019Natur.574..643R}. This suggests a rapid exhaustion of star formation fuel, at a level comparable to the star formation rate of several hundred M$_\odot$~yr$^{-1}$ \citep{2020ApJ...901..138P}. The ionized mass of either wind episode is unknown, however, and since no other gas phases have yet been observed in the extended Episode I wind, we do not know its outflow rate. In its time-resolved structure, Makani presents a unique opportunity to constrain the evolution of outflow rates with time.

Due to its compact size, most of the host galaxy itself is unresolved in ground-based observations. Based on Keck/NIRSPEC, MMT longslit, and SDSS data, we know the unresolved host emission---which includes a narrow component near systemic and a broad, blueshifted component that dominates the flux---lies in the AGN area of excitation diagrams \citep{2019Natur.574..643R,2021ApJ...923..275P}. The outflowing component in the galaxy also shows weak [\ion{Ne}{5}]~3324, 3426~\AA\ emission \citep{2019Natur.574..643R}. Normally these would be indicative of an AGN, but Makani shows no other evidence of an AGN \citep{2014MNRAS.441.3417S,2019Natur.574..643R}. We have tentatively interpreted these as evidence of high-velocity shocks as the Episode II wind propagates through the gas in the center of the galaxy. However, it is also possible that the ionization of the Makani host galaxy and that of its parent sample \citep{2007ApJ...663L..77T} reflect leakage of Lyman continuum (LyC) photons \citep{2021ApJ...923..275P}. Further constraining its ionization may distinguish between these scenarios.

To address these questions, here we present Keck/ESI long-slit spectra of Makani in the optical. In Section \ref{sec:obs}, we present the observations, data reduction, and data analysis. We describe the results in Section \ref{sec:results}, and discuss them in the light of shock and outflow models in Section~\ref{sec:discuss}. We summarize in the final Section~\ref{sec:conclude}.

\section{Observations and Data Processing} \label{sec:obs}

We observed Makani on 12 Sep 2020 UT with the Echellette Spectrograph and Imager (ESI; \citealt{2002PASP..114..851S}) on Keck II. We chose the 1\farcs0 slit in echellette mode, which yields moderate velocity resolution ($R = 4000$). The broad, simultaneous wavelength coverage allows us to observe many rest-frame optical strong lines ([\ion{O}{2}] through [\ion{S}{2}]~6716,~6731~\AA), while the 20\arcsec\ slit length is well-matched to the extent of the 17\arcsec\ nebula.

\begin{figure}
    \centering
    \includegraphics[width=\columnwidth]{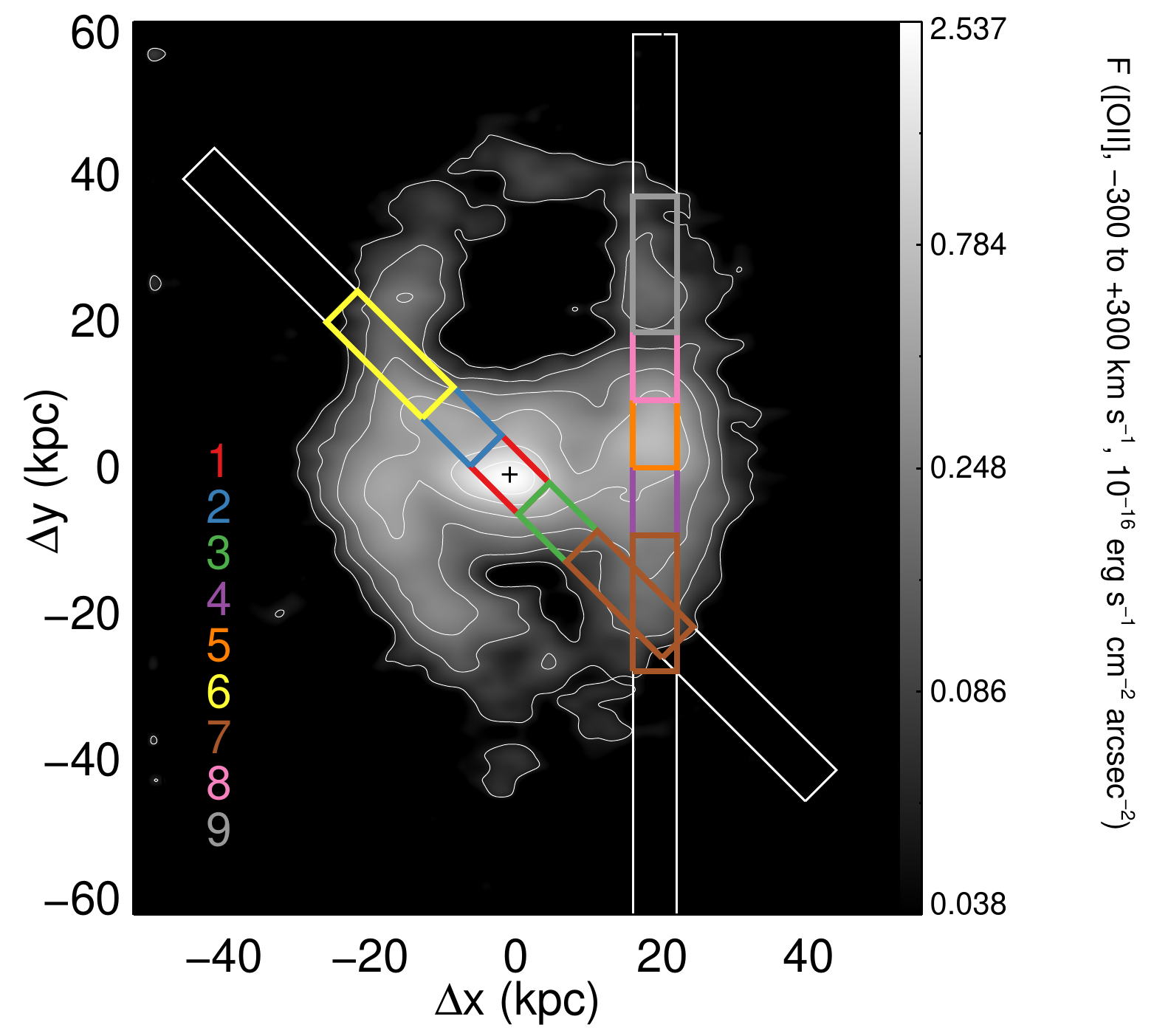}
    \caption{The Makani [\ion{O}{2}] nebula as observed by KCWI \citep{2019Natur.574..643R}, with ESI long slits overlaid. The peak [OII] KCWI spaxel is labeled with a cross. The extracted apertures are shown as colored boxes. North is up and East is left.}
    \label{fig:slits}
\end{figure}

Conditions were photometric, with seeing $\la$0\farcs8. We took 3$\times$30 minute exposures centered on the galaxy with $\mathrm{PA}=45^\circ$ E of N, and 4$\times$30 minute exposures centered on a point offset 3\farcs3 W at $\mathrm{PA}=0^\circ$ (Figure~\ref{fig:slits}). These slit positions were designed to probe the highest-surface brightness regions of the nebula outside of the nucleus while catching significant areas of low surface-brightness features. To calibrate, we took bias, arc lamp, dome flat, twilight flat, and pinhole flat observations. We dithered $\pm$1\arcsec\ along the slit between exposures. We observed the flux standard BD+28 4211 for photometric and telluric corrections. Airmasses were 1.06--1.07 for the flux standard and $\mathrm{PA}=45^\circ$ slit; and 1.1--1.4 for the $\mathrm{PA}=0^\circ$ slit.

We reduce the data using the ESIRedux package \citep{2003ApJS..147..227P} in XIDL. We use all calibrations but the twilight flats. We additionally telluric-correct the spectra using the flux standard after normalizing it by a 82,000 K blackbody. The final spectra are in vacuum wavelengths.

Along each slit, we extract five apertures (Figure \ref{fig:slits} and Table \ref{tab:aps}, labelled in order of increasing projected galactocentric radius). The lengths of the inner apertures are 2$\times$FWHM$_\mathrm{seeing}$ (10 pixels, or 1\farcs54), and those of the outer apertures are 4$\times$FWHM$_\mathrm{seeing}$ (20 pixels, or 3\farcs08). We coadd the two overlapping apertures (ap7). We searched for emission in apertures outside the previously detected nebula but found none that surpassed the detection threshold.

We extract a 3\arcsec$\times$1\arcsec\ rectangular aperture centered on the galaxy to compare to the existing 3\arcsec\ diameter SDSS spectrum. The SDSS spectrum is 26\%\ higher in flux over 6500--9000~\AA, but the difference rises smoothly with decreasing wavelength to 80\%\ at 4000~\AA. This flux difference is also observed in a 3\arcsec\ circular extraction from the KCWI data cube. We apply this upward correction to each extracted spectrum. The effect on flux ratios is minimal, but may be 5\%\ for [\ion{O}{2}]/H$\alpha$, as the total upward correction at [\ion{O}{2}] is 33\%.

\begin{deluxetable}{cccc}
  \tablecaption{Apertures\label{tab:aps}}
  \tablewidth{0pt}

  \tablehead{\colhead{Label} & \colhead{Size} & \colhead{Radius} & \colhead{f([\ion{O}{2}])/10$^{-16}$} \\
  \colhead{} & \colhead{} & \colhead{kpc} & \colhead{erg~s$^{-1}$~cm$^{-2}$} \\   
  \colhead{(1)} & \colhead{(2)}  & \colhead{(3)} & \colhead{(4)}}

  \startdata
ap1   &   1\arcsec$\times$1\farcs54 & 0.0$_{}^{5.5}$ & 7.74$\pm$0.26 \\
ap2   &   1\arcsec$\times$1\farcs54 & 9.3$_{3.8}^{5.0}$ & 2.30$\pm$0.08 \\
ap3   &   1\arcsec$\times$1\farcs54 & 9.3$_{3.8}^{5.0}$ & 1.50$\pm$0.11 \\
ap4   &   1\arcsec$\times$1\farcs54 & 20.2$_{3.3}^{4.1}$ & 0.75$\pm$0.03 \\
ap5   &   1\arcsec$\times$1\farcs54 & 20.6$_{3.8}^{4.4}$ & 1.41$\pm$0.03 \\
ap6   &   1\arcsec$\times$3\farcs08 & 23.2$_{9.0}^{9.4}$ & 0.94$\pm$0.06 \\
ap7   &   1\arcsec$\times$3\farcs08 & 23.2$_{9.0}^{9.4}$ & 0.39$\pm$0.05 \\
ap8   &   1\arcsec$\times$1\farcs54 & 24.8$_{5.1}^{5.3}$ & 0.45$\pm$0.02 \\
ap9   &   1\arcsec$\times$3\farcs08 & 35.0$_{9.2}^{9.4}$ & 0.19$\pm$0.02 \\
 \enddata

  \tablecomments{Column 2: Aperture extraction size. Column 3: Radius of the aperture center, plus or minus the full range of radii represented by the boundaries of the aperture. Column 4: Observed [\ion{O}{2}] 3726, 3729~\AA\ flux in the aperture, with 1$\sigma$ errors.}
  
\end{deluxetable}

\begin{figure*}
    \centering
    \includegraphics[width=\textwidth]{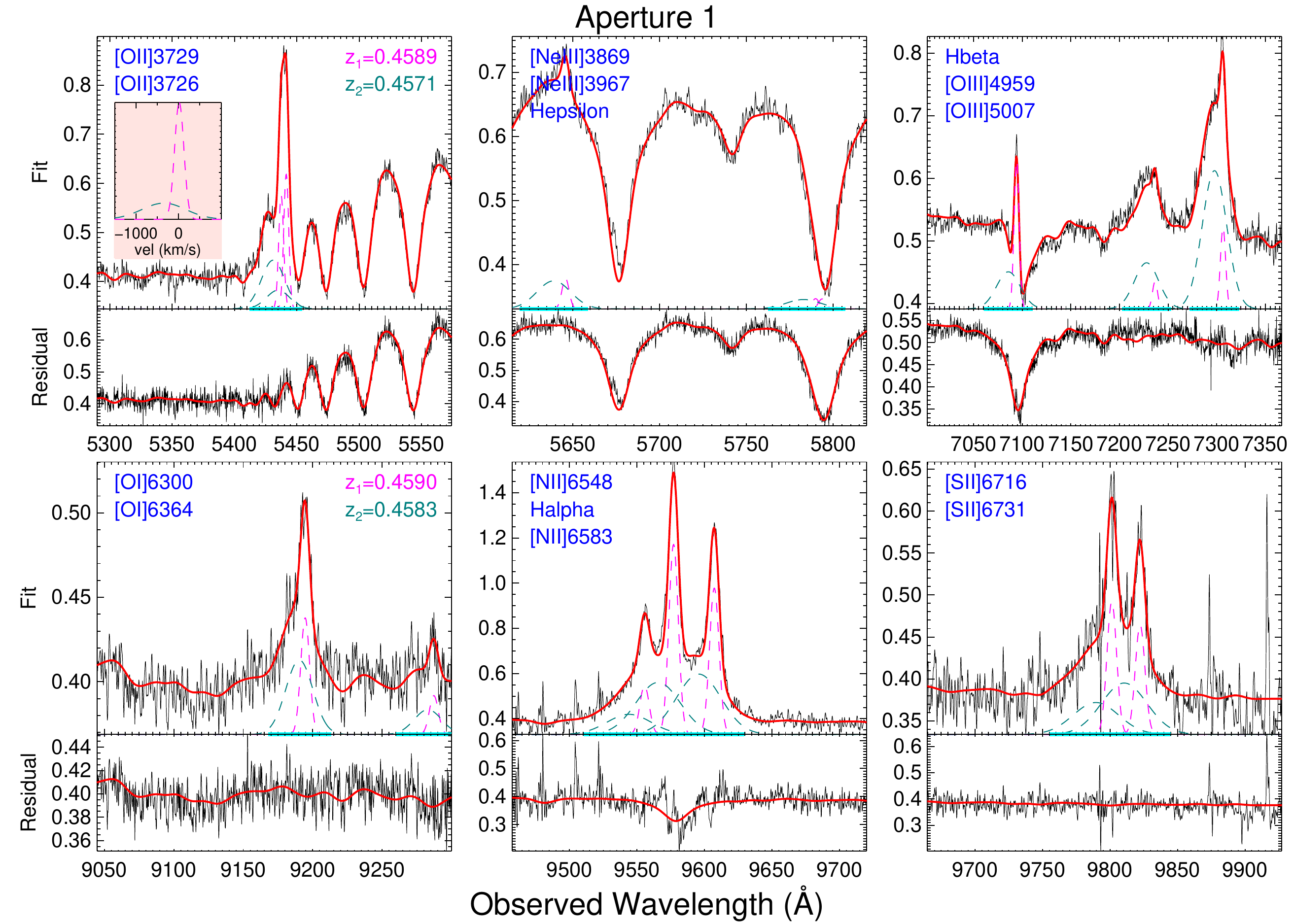}
    \caption{Observed spectrum of each aperture and model fits near strong emission lines. In each panel, the lines plotted are in the upper left-hand corner. The black line is the data, in units of 10$^{-16}$ erg~s$^{-1}$~cm$^{-2}$~\AA$^{-1}$; the red line is the total model fit (lines$+$continuum); and dashed lines are the emission-line model. Different velocity components are shown as different colors. The best-fit redshift of each component is shown in the [\ion{O}{2}] panel, as is an inset showing the velocity profile of [\ion{O}{2}]~3729~\AA. Because the velocity of [\ion{O}{1}] is allowed to vary separately in some apertures, its best-fit redshifts are shown in the corresponding panel. The data are smoothed with a 3-pixel boxcar. The data and model residuals after subtraction of the emission-line model (black and red lines) and the 1$\sigma$ error in the data (teal line) are shown in the bottom third of each panel. (The complete figure set (9 images) is available.)}
    \label{fig:spec}
\end{figure*}

We fit the spectra using IFSFIT \citep{2014ascl.soft09005R}. We fit the continuum with pPXF \citep{2012ascl.soft10002C,2017MNRAS.466..798C} and the same C3K Solar-metallicity stellar models (Conroy et al., in prep.) used in \citet{2019Natur.574..643R}. We use the best-fit stellar model to the host spectrum (ap1) as a template for the two $r\sim10$~kpc apertures (ap2 and ap3), which also contain some stellar emission. In these fits, we allow for non-zero stellar attenuation and include an order-4 additive polynomial to account for uncertainties in data reduction (flux calibration, scattered light, etc.). The only aperture with a non-zero best-fit stellar attenuation is ap1, for which $E(B-V)_* = 0.3$. There is no visible stellar continua in other apertures, though very low levels of continuum flux may indicate imperfect sky subtraction in some cases. To remove this residual signal, we fit an order-10 polynomial continuum.

After subtracting the best-fit continuum model, we fit two velocity components---one narrow and near systemic, the other broad and blueshifted---to the emission lines in ap1, ap2, and ap3, and one component otherwise (Figure~\ref{fig:spec}). (Hereafter we refer to the narrow, near-systemic component of ap1 as ap1.1 and the broad, blueshifted component of ap1 as ap1.2.) These components are robustly detected in the host galaxy and we treat them separately in some parts of the analysis. While multiple components (including the same broad line seen in ap1) are clearly required in ap2 and ap3, the lower S/N in some lines means that these components are not significantly detected in every line. For instance, [\ion{S}{2}] suffers from lower throughput and sky line contamination. Thus, in our analysis we treat only the total flux for ap2 and ap3. For these apertures, we also fix the [\ion{O}{2}] 3729/3726 flux ratios to match that of ap1.

For the most part, the velocities and linewidths of all emission lines are tied together. However, we find that the fit to [\ion{O}{1}] in apertures ap1, ap4, and ap5 is improved if we fit its velocity and linewidth separately. (In other apertures, the line is too faint to fit separately.) The most significant difference is that in ap1, the broad component in [\ion{O}{1}] is redshifted compared to the broad component in other lines by $\Delta v = 350$~km/s and narrower by $\Delta \sigma = 200$~km/s (40\%\ smaller). In the narrow component of ap1 and in ap4 and ap5, [\ion{O}{1}] is only slightly redshifted ($\la$50~km/s) and the linewidth difference is small ($\la$30~km/s).

In several apertures (ap2, ap7, and ap9) we first fit [\ion{O}{2}] only and use the resulting line centers and linewidths as fixed priors for the other lines. As this line is the brightest and highest S/N, this serves to improve the fit for faint lines ([\ion{O}{1}]), lines impacted by sky contamination (in the red), and lines with some degeneracy due to blending (H$\alpha$ and [\ion{N}{2}]).

In ap9, H$\alpha$ is only marginally detected, as it is impacted by sky lines. For this aperture, we instead estimate H$\alpha$ from H$\beta$ assuming no extinction and the Case B ratio of 2.86.

\begin{figure}
    \centering
    \includegraphics[width=\columnwidth]{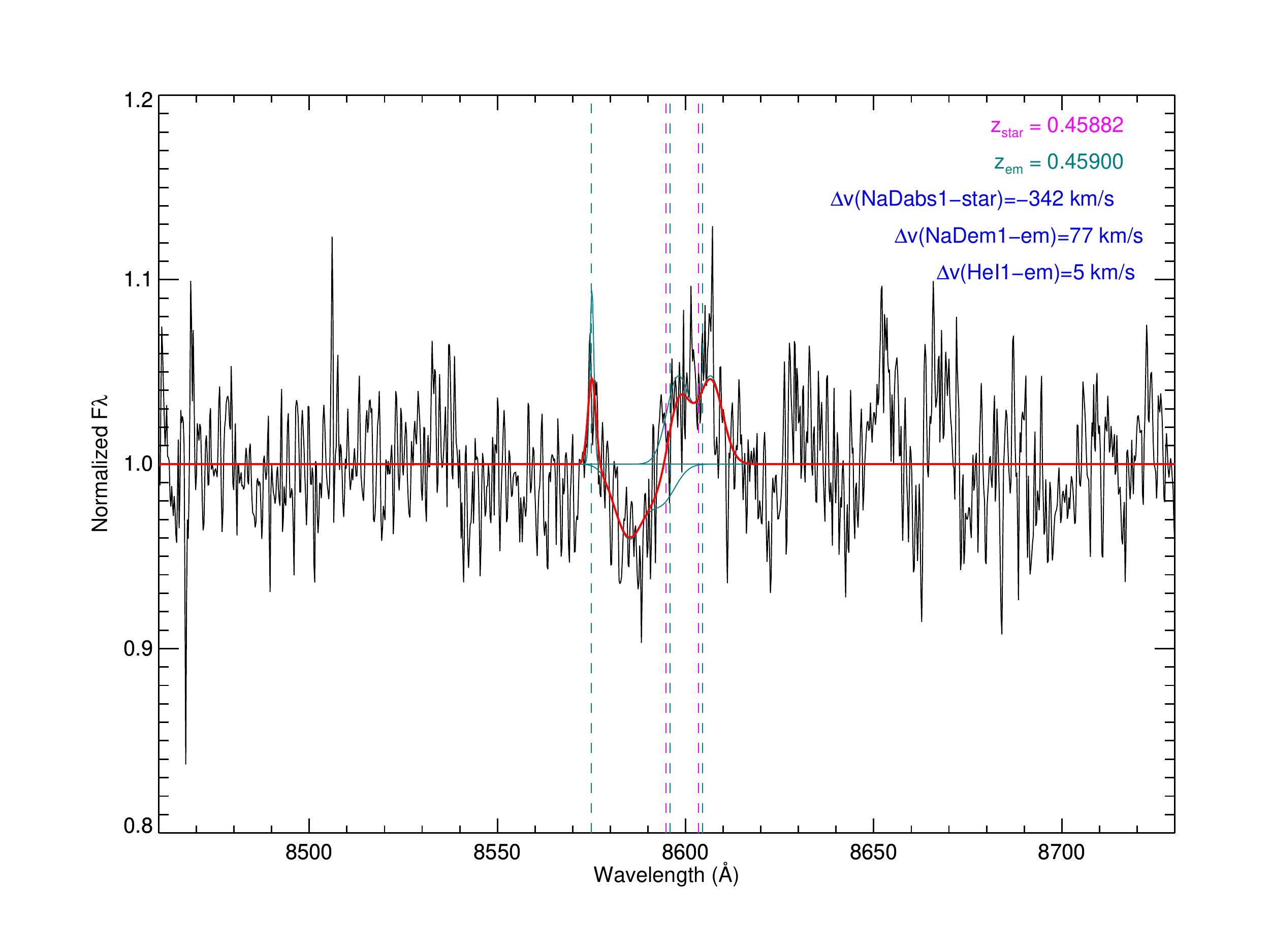}
    \caption{Fit to \ion{He}{1}~5876~\AA\ emission and \ion{Na}{1}~D~5890,~5896~\AA\ absorption and emission in the central spectrum, ap1. Vertical dashed lines show the expected wavelengths based on our fit to the stellar continuum (magenta) and [\ion{O}{1}] emission lines (teal) in ap1. Velocities with respect to these reference redshifts ($z_\mathrm{star}$ for \ion{Na}{1}~D and $z_\mathrm{em}$ for \ion{He}{1}) are shown in the upper right.}
    \label{fig:nad}
\end{figure}

There is residual interstellar \ion{Na}{1}~D absorption and emission in two apertures, ap1 and ap2. We fit these features also using IFSFIT (Figure~\ref{fig:nad}), and report the results in Section~\ref{sec:neutral}.

\section{Results} \label{sec:results}

The high throughput and long wavelength coverage of ESI allow us to detect numerous strong emission lines across the nebula. The ap1 spectrum also includes weak emission lines such as [\ion{Ne}{5}]~3426~\AA, [\ion{O}{3}]~4363~\AA, and [\ion{N}{2}]~5755~\AA. We report observed and extinction-corrected line fluxes of all emission lines detected at $>$2$\sigma$ in Table~\ref{tab:flux}, as well as H$\alpha$ fluxes and luminosities. Observed fluxes have been corrected for a Galactic extinction of $E(B-V)=0.0684$ \citep{2011ApJ...737..103S}.

\begin{deluxetable*}{cccccccccccc}
  \tablecaption{Fluxes and Luminosities\label{tab:flux}}
  \rotate
  \tabletypesize{\scriptsize}
  \tablewidth{0pt}
  \tablehead{\colhead{Line}& \colhead{ap1.1\tablenotemark{a}} & \colhead{ap1.2\tablenotemark{a}} & \colhead{ap1}  & \colhead{ap2} &
  \colhead{ap3} & \colhead{ap4} & \colhead{ap5} & \colhead{ap6} & \colhead{ap7} & 
  \colhead{ap8} & \colhead{ap9} \\
  \colhead{(1)} & \colhead{(2)}  & \colhead{(3)} & \colhead{(4)} & \colhead{(5)} &
  \colhead{(6)} & \colhead{(7)}  & \colhead{(8)} & \colhead{(9)} & \colhead{(10)} & 
  \colhead{(11)} & \colhead{(12)}
  }
  \startdata
           [NeV]3426 &          \nodata &   -1.43$\pm$0.19 &   -1.66$\pm$0.20 &          \nodata &          \nodata &          \nodata &          \nodata &          \nodata &          \nodata &          \nodata &          \nodata \\\relax
           [OII]3726 &   -0.74$\pm$0.03 &   -0.55$\pm$0.02 &   -0.63$\pm$0.02 &   -0.35$\pm$0.03 &   -0.33$\pm$0.05 &   -0.30$\pm$0.04 &   -0.31$\pm$0.02 &   -0.08$\pm$0.07 &   -0.27$\pm$0.06 &   -0.01$\pm$0.06 &   -0.05$\pm$0.09 \\\relax
           [OII]3729 &   -0.67$\pm$0.04 &   -0.97$\pm$0.05 &   -0.81$\pm$0.03 &   -0.45$\pm$0.03 &   -0.57$\pm$0.07 &   -0.14$\pm$0.04 &   -0.14$\pm$0.02 &    0.08$\pm$0.07 &   -0.42$\pm$0.13 &    0.15$\pm$0.06 &    0.11$\pm$0.07 \\\relax
 [OII]3726+[OII]3729 &   -0.41$\pm$0.03 &   -0.41$\pm$0.02 &   -0.41$\pm$0.02 &   -0.09$\pm$0.03 &   -0.13$\pm$0.05 &    0.09$\pm$0.03 &    0.09$\pm$0.02 &    0.31$\pm$0.07 &   -0.04$\pm$0.07 &    0.38$\pm$0.05 &    0.34$\pm$0.06 \\\relax
         [NeIII]3869 &   -1.42$\pm$0.08 &   -0.89$\pm$0.04 &   -1.05$\pm$0.03 &   -0.94$\pm$0.09 &          \nodata &          \nodata &   -1.12$\pm$0.07 &   -0.58$\pm$0.13 &   -0.88$\pm$0.11 &   -0.77$\pm$0.15 &          \nodata \\\relax
          [OIII]4363 &   -1.86$\pm$0.20 &   -1.37$\pm$0.11 &   -1.52$\pm$0.09 &   -1.05$\pm$0.12 &          \nodata &          \nodata &          \nodata &          \nodata &          \nodata &          \nodata &          \nodata \\\relax
               H$\beta$ &   -0.65$\pm$0.01 &   -0.69$\pm$0.02 &   -0.67$\pm$0.01 &   -0.87$\pm$0.06 &   -0.58$\pm$0.06 &   -0.66$\pm$0.05 &   -0.64$\pm$0.03 &   -0.68$\pm$0.13 &   -0.74$\pm$0.11 &   -0.41$\pm$0.07 &   -0.46$\pm$0.22 \\\relax
          [OIII]5007 &   -0.90$\pm$0.03 &   -0.11$\pm$0.02 &   -0.31$\pm$0.01 &   -0.48$\pm$0.04 &   -0.61$\pm$0.06 &   -0.67$\pm$0.07 &   -0.74$\pm$0.04 &   -0.44$\pm$0.11 &          \nodata &   -0.60$\pm$0.11 &   -0.30$\pm$0.21 \\\relax
   [NI]5198+[NI]5200 &   -1.11$\pm$0.07 &          \nodata &   -1.39$\pm$0.08 &   -1.61$\pm$0.09 &   -1.40$\pm$0.11 &   -1.43$\pm$0.11 &          \nodata &          \nodata &          \nodata &          \nodata &          \nodata \\\relax
           [NII]5755 &   -1.66$\pm$0.11 &   -1.33$\pm$0.09 &   -1.44$\pm$0.07 &          \nodata &          \nodata &          \nodata &          \nodata &          \nodata &          \nodata &          \nodata &          \nodata \\\relax
            [OI]6300 &   -1.12$\pm$0.08 &   -0.94$\pm$0.09 &   -1.01$\pm$0.06 &   -0.73$\pm$0.05 &   -0.70$\pm$0.07 &   -0.68$\pm$0.08 &   -0.69$\pm$0.04 &          \nodata &          \nodata &   -0.57$\pm$0.09 &   -0.32$\pm$0.15 \\\relax
           [NII]6583 &   -0.11$\pm$0.02 &    0.07$\pm$0.03 &   -0.00$\pm$0.02 &   -0.20$\pm$0.06 &   -0.31$\pm$0.10 &   -0.46$\pm$0.08 &   -0.48$\pm$0.04 &   -0.20$\pm$0.14 &   -0.60$\pm$0.14 &          \nodata &   -0.06$\pm$0.07 \\\relax
           [SII]6716 &   -0.72$\pm$0.03 &   -0.77$\pm$0.06 &   -0.75$\pm$0.03 &   -0.54$\pm$0.06 &   -0.24$\pm$0.05 &   -0.17$\pm$0.05 &   -0.34$\pm$0.03 &          \nodata &    0.04$\pm$0.04 &   -0.90$\pm$0.15 &          \nodata \\\relax
           [SII]6731 &   -0.80$\pm$0.05 &   -0.57$\pm$0.11 &   -0.65$\pm$0.08 &   -0.61$\pm$0.10 &   -0.41$\pm$0.06 &   -0.33$\pm$0.06 &   -0.50$\pm$0.04 &   -0.73$\pm$0.14 &   -0.12$\pm$0.05 &   -0.55$\pm$0.08 &          \nodata \\\relax
 [SII]6716+[SII]6731 &   -0.46$\pm$0.03 &   -0.36$\pm$0.07 &   -0.40$\pm$0.05 &   -0.28$\pm$0.06 &   -0.02$\pm$0.05 &    0.06$\pm$0.04 &   -0.11$\pm$0.02 &   -0.57$\pm$0.14 &    0.27$\pm$0.04 &   -0.39$\pm$0.08 &          \nodata \\
\hline
                log($F_\mathrm{H\alpha}$) & -15.063$\pm$0.009 & -14.952$\pm$0.014 & -14.703$\pm$0.009 & -15.55$\pm$0.02 & -15.69$\pm$0.04 & -16.21$\pm$0.03 & -15.94$\pm$0.01 & -16.34$\pm$0.06 & -16.37$\pm$0.03 & -16.72$\pm$0.05 & -17.06$\pm$0.05 \\
\hline
                        E(B-V) &   0.454$\pm$0.035 &   0.546$\pm$0.056 &   0.505$\pm$0.034 &   0.97$\pm$0.14 &   0.30$\pm$0.14 &   0.48$\pm$0.12 &   0.42$\pm$0.06 &   0.51$\pm$0.30 &   0.67$\pm$0.26 &   0.00$\pm$0.17 &   0.00$\pm$0.51 \\
\hline
           [NeV]3426 &          \nodata &   -0.89$\pm$0.20 &   -1.17$\pm$0.21 &          \nodata &          \nodata &          \nodata &          \nodata &          \nodata &          \nodata &          \nodata &          \nodata \\\relax
           [OII]3726 &   -0.34$\pm$0.04 &   -0.06$\pm$0.06 &   -0.17$\pm$0.04 &    0.52$\pm$0.14 &   -0.06$\pm$0.14 &    0.13$\pm$0.12 &    0.07$\pm$0.06 &    0.38$\pm$0.29 &    0.33$\pm$0.25 &   -0.01$\pm$0.17 &   -0.05$\pm$0.48 \\\relax
           [OII]3729 &   -0.27$\pm$0.05 &   -0.48$\pm$0.07 &   -0.36$\pm$0.04 &    0.42$\pm$0.14 &   -0.31$\pm$0.15 &    0.29$\pm$0.12 &    0.23$\pm$0.06 &    0.54$\pm$0.29 &    0.18$\pm$0.27 &    0.15$\pm$0.17 &    0.11$\pm$0.48 \\\relax
 [OII]3726+[OII]3729 &    0.00$\pm$0.04 &    0.08$\pm$0.06 &    0.04$\pm$0.04 &    0.77$\pm$0.14 &    0.14$\pm$0.14 &    0.52$\pm$0.12 &    0.46$\pm$0.06 &    0.77$\pm$0.29 &    0.56$\pm$0.25 &    0.38$\pm$0.17 &    0.34$\pm$0.48 \\\relax
         [NeIII]3869 &   -1.03$\pm$0.08 &   -0.43$\pm$0.07 &   -0.62$\pm$0.05 &   -0.12$\pm$0.16 &          \nodata &          \nodata &   -0.77$\pm$0.09 &   -0.14$\pm$0.31 &   -0.31$\pm$0.26 &   -0.77$\pm$0.21 &          \nodata \\\relax
          [OIII]4363 &   -1.57$\pm$0.20 &   -1.02$\pm$0.12 &   -1.20$\pm$0.10 &   -0.43$\pm$0.17 &          \nodata &          \nodata &          \nodata &          \nodata &          \nodata &          \nodata &          \nodata \\\relax
               H$\beta$ &   -0.46$\pm$0.03 &   -0.46$\pm$0.05 &   -0.46$\pm$0.03 &   -0.46$\pm$0.12 &   -0.46$\pm$0.12 &   -0.46$\pm$0.11 &   -0.46$\pm$0.05 &   -0.46$\pm$0.26 &   -0.46$\pm$0.22 &   -0.41$\pm$0.15 &   -0.46$\pm$0.44 \\\relax
          [OIII]5007 &   -0.73$\pm$0.04 &    0.09$\pm$0.04 &   -0.12$\pm$0.03 &   -0.12$\pm$0.11 &   -0.49$\pm$0.12 &   -0.49$\pm$0.11 &   -0.58$\pm$0.06 &   -0.25$\pm$0.25 &          \nodata &   -0.60$\pm$0.17 &   -0.30$\pm$0.43 \\\relax
   [NI]5198+[NI]5200 &   -0.97$\pm$0.07 &          \nodata &   -1.23$\pm$0.09 &   -1.31$\pm$0.13 &   -1.30$\pm$0.15 &   -1.28$\pm$0.14 &          \nodata &          \nodata &          \nodata &          \nodata &          \nodata \\\relax
           [NII]5755 &   -1.58$\pm$0.11 &   -1.24$\pm$0.10 &   -1.36$\pm$0.07 &          \nodata &          \nodata &          \nodata &          \nodata &          \nodata &          \nodata &          \nodata &          \nodata \\\relax
            [OI]6300 &   -1.10$\pm$0.08 &   -0.91$\pm$0.09 &   -0.99$\pm$0.07 &   -0.68$\pm$0.10 &   -0.68$\pm$0.11 &   -0.65$\pm$0.11 &   -0.67$\pm$0.05 &          \nodata &          \nodata &   -0.57$\pm$0.14 &   -0.32$\pm$0.36 \\\relax
           [NII]6583 &   -0.11$\pm$0.03 &    0.06$\pm$0.04 &   -0.01$\pm$0.03 &   -0.20$\pm$0.11 &   -0.31$\pm$0.13 &   -0.46$\pm$0.11 &   -0.48$\pm$0.05 &   -0.20$\pm$0.23 &   -0.60$\pm$0.21 &          \nodata &   -0.06$\pm$0.32 \\\relax
           [SII]6716 &   -0.73$\pm$0.04 &   -0.79$\pm$0.07 &   -0.76$\pm$0.04 &   -0.57$\pm$0.11 &   -0.25$\pm$0.10 &   -0.19$\pm$0.09 &   -0.35$\pm$0.05 &          \nodata &    0.02$\pm$0.16 &   -0.90$\pm$0.18 &          \nodata \\\relax
           [SII]6731 &   -0.82$\pm$0.06 &   -0.58$\pm$0.11 &   -0.67$\pm$0.08 &   -0.64$\pm$0.13 &   -0.41$\pm$0.10 &   -0.35$\pm$0.10 &   -0.52$\pm$0.05 &   -0.75$\pm$0.23 &   -0.14$\pm$0.17 &   -0.55$\pm$0.13 &          \nodata \\\relax
 [SII]6716+[SII]6731 &   -0.47$\pm$0.04 &   -0.37$\pm$0.08 &   -0.41$\pm$0.05 &   -0.31$\pm$0.10 &   -0.03$\pm$0.10 &    0.04$\pm$0.09 &   -0.13$\pm$0.04 &   -0.59$\pm$0.23 &    0.25$\pm$0.16 &   -0.39$\pm$0.13 &          \nodata \\
\hline
      log($L_\mathrm{H\alpha}$) & 42.32$^{+0.02}_{-0.02}$ & 42.52$^{+0.03}_{-0.03}$ & 42.73$^{+0.02}_{-0.02}$ & 42.36$^{+0.06}_{-0.07}$ & 41.53$^{+0.07}_{-0.08}$ & 41.20$^{+0.06}_{-0.07}$ & 41.41$^{+0.03}_{-0.03}$ & 41.10$^{+0.13}_{-0.18}$ & 41.22$^{+0.10}_{-0.14}$ & 40.20$^{+0.08}_{-0.10}$ & 39.86$^{+0.18}_{-0.32}$
  \enddata
  \tablecomments{Column 1: Emission line. Columns 2--12, top half: log of the observed line flux, normalized to H$\alpha$. Fluxes have been corrected for Galactic extinction. Following are the log of H$\alpha$ fluxes in erg~s$^{-1}$~cm$^{-2}$. Errors are 1$\sigma$ in the log. Middle: Color excess, computed from H$\alpha$/H$\beta$ for Case B conditions using the \citet{1989ApJ...345..245C} extinction curve and $R_V = 3.1$. Bottom half: log of the extinction-corrected line flux, normalized to H$\alpha$. Following are log of H$\alpha$ luminosities in erg~s$^{-1}$.}
  \tablenotetext{a}{ap1.1 and ap1.2 refer to the narrow, near-systemic and broad, blueshifted emission-line components, respectively.}
\end{deluxetable*}

\begin{figure}
    \centering
    \includegraphics[width=\columnwidth]{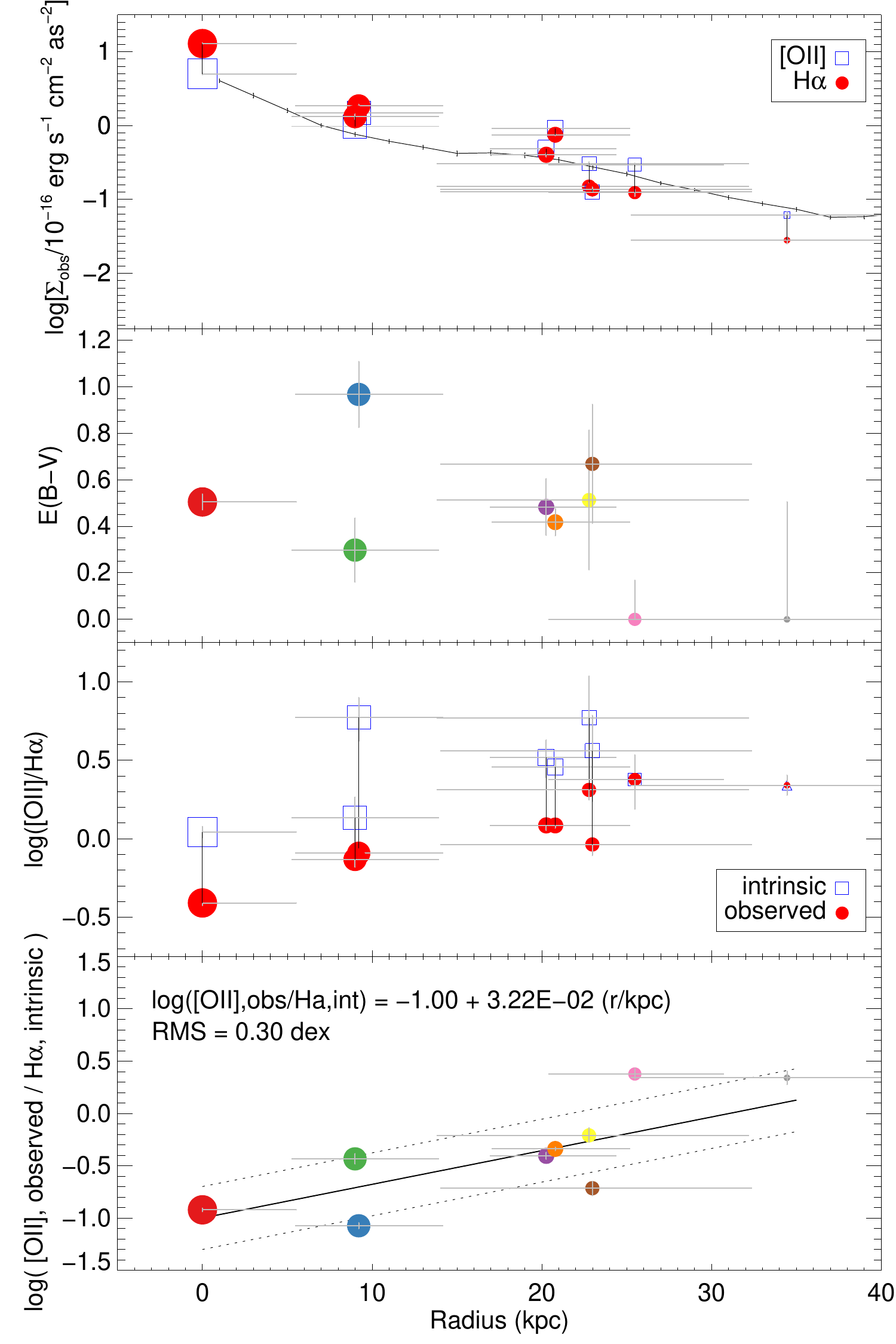}
    \caption{(Top) Radial dependence of observed [\ion{O}{2}] and H$\alpha$ surface brightness in each spectrum. The line shows the azimuthally-averaged radial surface brightness profile of [\ion{O}{2}] measured with KCWI \citep{2019Natur.574..643R}. Each point is centered on the radius of the aperture center; horizontal grey lines are the radii spanned by each extraction aperture; and vertical grey lines are 1$\sigma$ errors. Points in the same aperture are connected by vertical bars, and points at radius $>$ 0~kpc are randomly offset slightly in radius to prevent visual overlap. (Upper middle) Color excess $E(B-V)$ vs. radius. $E(B-V)$ is computed from the Balmer decrement. Vertical error bars are 1$\sigma$. Point colors correpond to the aperture colors from Figure \ref{fig:slits}. Reddening is observed out to 20~kpc. (Lower middle) Variation of observed and intrinsic [\ion{O}{2}]/H$\alpha$] flux ratio with distance from the host galaxy. The final point is a lower limit due to the large uncertainty in $E(B-V)$. Both ratios increase steadily from the host to the outskirts of the nebula. (Bottom) Correction from observed [\ion{O}{2}] to intrinsic H$\alpha$ vs. radius. The lines show a linear fit to the data and (fit$\pm$RMS).}
    \label{fig:o2ha}
\end{figure}

The ESI spectra cover the full range of Balmer lines. In Figure \ref{fig:o2ha}, we show observed [\ion{O}{2}] and H$\alpha$ surface brightnesses as a function of radius. The ratio steadily rises from about 1/3 in the center, to unity at 20 kpc, and finally to a maximum of 2 at $r = 30-40$~kpc. This increase in observed [\ion{O}{2}]/H$\alpha$ is due to a combination of decreasing extinction and increasing collisional excitation with increasing radius. We compute $E(B-V)$ from H$\alpha$/H$\beta$ for Case B conditions using the \citet{1989ApJ...345..245C} extinction curve and $R_V = 3.1$. $E(B-V)$ decreases with increasing radius from a peak of 1.0 at $r = 10$~kpc to 0 at $r>25$~kpc (Figure \ref{fig:o2ha}). The intrinsic [\ion{O}{2}]/H$\alpha$ flux ratio is near unity in ap1 and ap3, rising to a value of 2--5 in larger-radius apertures.

By combining the ESI H$\alpha$ and KCWI [\ion{O}{2}] measurements, we can estimate the spatially-resolved and integrated H$\alpha$ luminosity. We do so by bootstrapping from the ESI measurements to estimate how the ratio ([\ion{O}{2}], observed)/(H$\alpha$, intrinsic) changes with radius. We fit a linear relationship between 0 and 35 kpc (Figure~\ref{fig:o2ha}) with an RMS of 0.3~dex. We then apply this relationship to the Voronoi-binned KCWI data to infer the spatially-resolved H$\alpha$ flux. The resulting total H$\alpha$ luminosity is $L_\mathrm{H\alpha}^\mathrm{tot} = (9.4^{+6.0}_{-3.0})\times10^{42}$~erg~s$^{-1}$, where we calculate the error by applying the RMS dispersion at $r>5.5$~kpc (the edge of ap1). (Using instead a parabolic fit increases the luminosity by 15\%, but does not change the RMS and results in line ratios at the edge of the wind---$r\sim50$~kpc---that appear unphysical.)

We calculate the electron density using the flux ratios [\ion{O}{2}]3729/3726 and [\ion{S}{2}]6716/6731, which are allowed to freely vary between physical limits \citep{2016ApJ...816...23S}. Both lines are observed throughout the nebula, though [\ion{S}{2}] is not detected in all apertures. The [\ion{O}{2}] doublet is one of the brightest lines in the rest-frame optical spectrum of the nebula and is unaffected by sky lines. However, the lines are separated by only 3~\AA, leading to covariance among fit parameters. Overlap of multiple velocity components adds further degeneracy. While the [\ion{S}{2}] lines are better separated in wavelength (15~\AA), they are impacted by sky lines and fall in a lower-sensitivity part of the data. Thus, we report density estimates only in ap1, ap4, and ap5; the significance of the individual line detections of the [\ion{S}{2}] doublet, and thus our ability to reliably decompose them, is too low in other apertures, even if the total doublet is detected.

In ap1.1, densities are log($n_\mathrm{e}$/cm$^{-3}$) = 2.3$_{-0.6}^{+0.3}$ and 2.3$_{-0.3}^{+0.2}$ from [\ion{S}{2}] and [\ion{O}{2}], respectively. The ap1.2 (higher $\sigma$, blueshifted) densities are 3.4$_{-0.4}^{+0.6}$ and $>$4.0$_{-0.9}$. For ap4 and ap5, log($n_\mathrm{e}$/cm$^{-3}$) $<$ 1 from both line ratios, in both apertures. The upper error bar is rather large (1.5--2.3 dex) on the ap4--ap5 measurements, however, due to the decreasing sensitivity  of the line ratios to densities below $\sim$50~cm$^{-3}$.

\subsection{Excitation of the wind} \label{sec:excitation}

\begin{figure}
    \centering
    \includegraphics[width=\columnwidth]{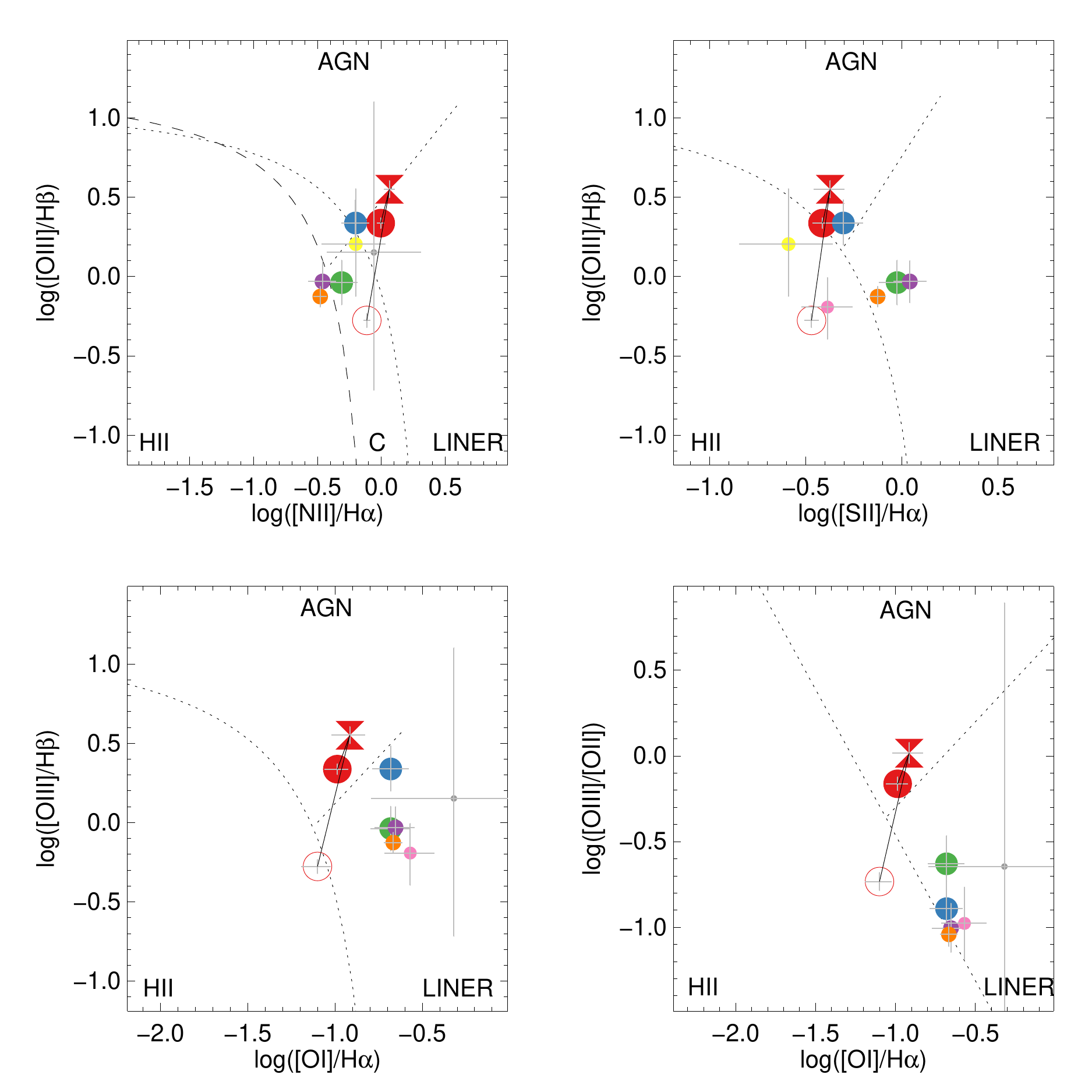}
    \caption{Excitation plots for the Makani nebula. The dashed and dotted lines divide nuclear classification regions of star-forming (\ion{H}{2}), AGN, LINER, and composite (C) using canonical dividing lines \citep{2003MNRAS.346.1055K,2006MNRAS.372..961K,2010MNRAS.403.1036C}. The symbol size reflects projected distance from the host galaxy, with smaller symbols at larger radius. Filled circles are from total fluxes. The bowtie is the central aperture broad component (ap1.2), and the open circle is the corresponding narrow component (ap1.1).}
    \label{fig:vo}
\end{figure}

The detection of multiple emission lines throughout the Makani nebula allows us to constrain the spatially-resolved excitation of the wind using standard line ratios (Figure~\ref{fig:vo}; \citealt{1981PASP...93....5B,1987ApJS...63..295V,2010MNRAS.403.1036C}).

Beginning with the nucleus, ap1, we that find that the broad component dominates its line flux and is securely in the AGN region of these diagrams. This is consistent with previous results \citep{2021ApJ...923..275P}. We extend this conclusion to the [\ion{O}{1}]/H$\alpha$ vs.\ [\ion{O}{3}]/H$\beta$ and [\ion{O}{1}]/H$\alpha$ vs.\ [\ion{O}{3}]/[\ion{O}{2}] diagrams, based on our detection of [\ion{O}{1}]. \citet{2021ApJ...923..275P} interpreted these line ratios as ionization from fast shocks; we discuss this in detail in Section~\ref{sec:discuss}.  Makani lacks of evidence for an AGN in the rest-frame UV, X-ray, radio, or infrared \citep{2012ApJ...755L..26D, 2014MNRAS.441.3417S, 2020ApJ...901..138P, 2022arXiv220913632W}. The rest-frame optical-UV continuum of Makani shows no indication of non-stellar emission, and a stellar population fit to the SED is perfectly consistent with the data \citep{2019Natur.574..643R}. Though [\ion{Ne}{5}] emission is present, its luminosity is low compared to AGN \citep{2019Natur.574..643R}. We thus discount AGN photoionization as the origin of these line ratios.

The narrow component of ap1 is instead likely ionized by stars, with some possible contribution from shocks based on its location in the composite region of the [\ion{N}{2}]/H$\alpha$ vs.\ [\ion{O}{3}]/H$\beta$ diagram.

The two apertures with $r\sim10$~kpc (ap2 and ap3) combine information from two velocity components, one broad and one narrow. Ap2 has line ratios similar to those from total fluxes in ap1, with higher low-ionization line ratios ([\ion{O}{1}]/H$\alpha$) and lower ionization parameter [\ion{O}{3}]/[\ion{O}{2}]. Apertures ap3, ap4, and ap5 have similar [\ion{O}{1}]/H$\alpha$ and [\ion{O}{3}]/[\ion{O}{2}] as ap2, but with lower high-ionization line ratios ([\ion{O}{3}]/H$\beta$), larger [\ion{S}{2}]/H$\alpha$, and slightly smaller [\ion{N}{2}]/H$\alpha$. Two other apertures with significant detections across several lines (ap6 and ap8) lie in roughly the same locations as the other apertures, but these apertures have low S/N in the weakest lines so their positioning is more uncertain. The lines other than [\ion{O}{2}] and H$\alpha$ in ap7 and ap9 are not strong enough to accurately place them in these diagrams, though ap9 does appear in several with large error bars.

Taken as a whole, the apertures other than ap1 lie largely in the composite region of the [\ion{N}{2}]/H$\alpha$ diagram and in the low-ionization nuclear emission-line region (LINER) area in the other three diagrams. An exception is aperture ap2, which lies in between the properties of ap1 and ap3. This aperture is in the direction of the most highly blueshifted molecular and ionized gas \citep{2019Natur.574..643R}. It must therefore include more flux from this higher-ionization wind component (mixed in with a lower-ionization component) than ap3, which is at the same radius as ap2.

We detect the weak emission lines [\ion{O}{3}]~4363~\AA\ and [\ion{N}{2}]~5755~\AA\ in ap1, and [\ion{O}{3}]~4363~\AA\ in ap2. The [\ion{O}{3}]~4363/5007 line flux ratios in ap1 and ap1.2 are -1.1~dex. Ratios this high are observed in LINERs, implying high temperatures ($T\ga2\times10^4$~K) \citep{2001ApJ...549..155N,2018ApJ...864...90M}. Similarly, the ratio [\ion{N}{2}]~5755/6548 is in the range -1.3 to -1.4, also consistent with $T\sim2\times10^4$~K.

\subsection{Neutral Wind Properties} \label{sec:neutral}

Makani may be the highest-redshift system in which interstellar \ion{Na}{1}~D absorption and/or emission has been seen in a galactic wind. Resonant \ion{Na}{1}~D absorption is detected in the host galaxy (ap1) with a low equivalent width (rest-frame $W_\mathrm{eq}$ of 0.37$^{+0.07}_{-0.05}$~\AA), but blueshifted by -342~km~s$^{-1}$ (Figure~\ref{fig:nad}). The outflow was not previously detected in resonant absorption in \ion{Mg}{2}~2796,~2803~\AA\ or \ion{Mg}{1} transitions in the near-UV \citep{2019Natur.574..643R}. However, there is residual, weak \ion{Fe}{2}~2585~\AA\ absorption \citep{2019Natur.574..643R,2021ApJ...923..275P}.

Previously, redshifted resonant emission in \ion{Mg}{2}, \ion{Mg}{1}, and \ion{Fe}{2}$^*$ was detected in Makani \citep{2019Natur.574..643R}. The \ion{Mg}{2} is spatially resolved into an $r\sim20$~kpc nebula. We now detect spatially-extended \ion{Na}{1}~D emission in ap1 and ap2, with fluxes and rest-frame equivalent  widths of $(3.2\pm0.2)\times10^{-17}$~erg~s$^{-1}$~cm$^{-2}$ and $W_\mathrm{eq}=-0.49\pm0.05$~\AA\ in ap1 and $(2.0\pm0.2)\times10^{-17}$~erg~s$^{-1}$~cm$^{-2}$ and $W_\mathrm{eq}=-4.4\pm0.5$~\AA\ in ap2.

Several lines of evidence suggest that resonant absorption in ap1 is filling in the emission, reducing the observed $|W_\mathrm{eq}|$ and increasing $|v-v_\mathrm{sys}|$ of both absorption and emission. First, a fit to \ion{Fe}{2}~2585~\AA\ gives a velocity at maximum depth of -100~km~s$^{-1}$ and a covering factor of 0.12, assuming it is optically thick. Thus $v_\mathrm{NaI}-v_\mathrm{FeII} \sim -200$~km~s$^{-1}$. Furthermore, if the \ion{Na}{1}~D and \ion{Fe}{2} absorption lines are optically thick, the ratio of the observed covering factors is $C_\mathrm{f,FeII}/C_\mathrm{f,NaI} \sim 3-4$, suggesting that \ion{Na}{1}~D is too shallow. 

Second, neutral gas absorption in outflows is closely connected with the foreground dust column \citep{2015ApJ...801..126R,2021MNRAS.503.4748R}. Though there is significant scatter, the observed $E(B-V)=0.6-1.0$ in ap1 and ap2 (Figure~\ref{fig:o2ha}) are typically associated with 5--10$\times$ higher $W_\mathrm{eq}$ on average.

Finally, a comparison to the radial surface-brightness profile of \ion{Mg}{2} (Figure~\ref{fig:neutral}) shows that the \ion{Na}{1}~D surface brightness declines much more slowly than that of \ion{Mg}{2}. If this is due to infilling of emission by foreground absorption, \ion{Na}{1}~D should intrinsically be brighter in ap1 by a factor of a few ($0.6\pm0.2$~dex). This in turn would raise the absorption equivalent width by the same factor and the covering factor by some (smaller) amount. This reduction of \ion{Na}{1}~D flux due to infilling is observed in other spatially-resolved observations \citep{2015ApJ...801..126R}.

Due to this infilling, we do not attempt to estimate the column density, and thus the mass and dynamics, of the neutral phase of the wind.

\begin{figure}
    \centering
    \includegraphics[width=\columnwidth]{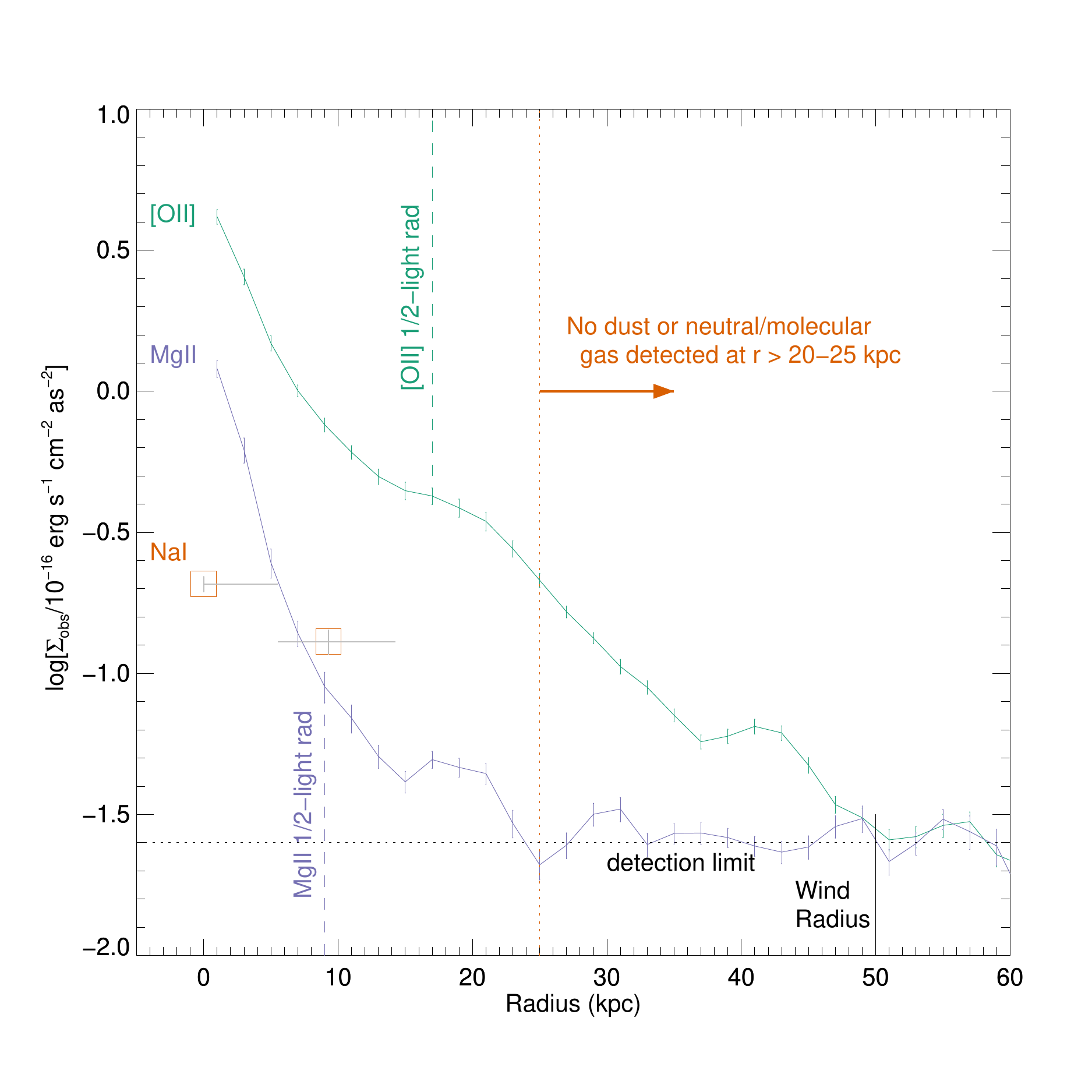}
    \caption{Radial variation of neutral gas tracers \ion{Mg}{2}~2796,~2803~\AA\ and \ion{Na}{1}~D in Makani. The blue line shows the azimuthally-averaged radial dependence of \ion{Mg}{2} surface brightness from KCWI data \citep{2019Natur.574..643R}, with half-light radius $r_\mathrm{e}^\mathrm{Mg~II} = 9$~kpc and maximum extent 20--25~kpc. The slower decline of \ion{Na}{1}~D emission, based on the current ESI data, suggests emission-line infilling in the innermost region, reducing its flux by $\sim$0.6~dex (Section~\ref{sec:neutral}). The lack of neutral/molecular gas (based on \ion{Mg}{2} and CO from \citealt{2019Natur.574..643R} and \ion{Na}{1} from the current work) and dust (based on $E(B-V)$ from the current work; Figure~\ref{fig:o2ha}) beyond 20--25~kpc is indicated with the vertical, orange dotted line. The full extent of the ionized gas nebula from KCWI observations is shown as the solid green line, with $r_\mathrm{e}^\mathrm{[O~II]} = 17$~kpc and $r_\mathrm{max}^\mathrm{[O~II]} = 50$~kpc.}
    \label{fig:neutral}
\end{figure}

\section{Discussion} \label{sec:discuss}

\subsection{Properties of the Star-Forming Host} \label{sec:starburst}

Makani contains a compact ($r\sim400$~pc) stellar core \citep{2014MNRAS.441.3417S, 2019Natur.574..643R, 2021ApJ...912...11D} with a young stellar population \citep{2019Natur.574..643R}. The low-velocity component of the innermost ($r\la5.5$~kpc) aperture, ap1.1, thus contains most or all of the star formation activity. Based on the observed line ratios (Section~\ref{sec:excitation} and \citealt{2021ApJ...923..275P}), this component is photoionized by the young stellar population. The H$\alpha$ luminosity of this component is $2.1\times10^{42}$~erg~s$^{-1}$ (Table~\ref{tab:flux}), which is 40\%\ of the total luminosity in ap1 (i.e., the host galaxy). We note that this luminosity, however, is larger than the difference between the total nebular luminosity (Section~\ref{sec:results}) and wind luminosity (Section~\ref{sec:mpe}). This difference is $L_\mathrm{H\alpha}^\mathrm{starburst} = L_\mathrm{H\alpha}^\mathrm{tot} - L_\mathrm{H\alpha}^\mathrm{wind} = (9.4-8.0)\times10^{42}$~erg~s$^{-1} = 1.4\times10^{42}$~erg~s$^{-1}$, 30\%\ lower than the value measured from ap1.1. We take this difference as an indication of the  uncertainties in our methodology, which includes assigning fluxes among the starburst and wind Episodes I and II, bootstrapping from the KCWI [\ion{O}{2}] data from the ESI meaurements, and the absolute flux calibration of the current data (Section~\ref{sec:obs}). In what follows we assume the average of these two values, $L_\mathrm{H\alpha}^\mathrm{starburst} = 1.8\times10^{42}$~erg~s$^{-1}$.

Makani has a star formation rate (SFR) of 224--300~M$_\odot$~yr$^{-1}$, as inferred from radio and infrared data \citep{2020ApJ...901..138P}.  The radio emission is concentrated in a 1\arcsec\ point source coinciding with ap1. Using a standard calibration \citep{2006ApJ...642..775M}, we compute from $L_\mathrm{H\alpha}^\mathrm{starburst}$ a SFR of 14~M$_\odot$~yr$^{-1}$. This is a factor of 16 smaller than the lowest previous estimate from IR data. A similar discrepancy is seen in other galaxies in the parent galaxy sample of Makani \citep{2022AAS...24024150E}. It may in fact be evidence of the stellar feedback in Makani shutting off star formation, as H$\alpha$ traces star formation over the shortest timescales $\la$6~Myr \citep{2013seg..book..419C}. This is just below the estimated 7~Myr estimate of the most recent burst of star formation. An earlier burst occurred 400~Myr ago \citep{2019Natur.574..643R}, which is unlikely to be reflected in current SFR estimates. The observed discrepancy could also result in part from LyC leakage \citep{2006ApJ...642..775M}, to which compact starbursts with strong feedback like Makani may be susceptible \citep{2021ApJ...923..275P}.

The bright, extended \ion{Mg}{2} emission in Makani is consistent with some LyC escape \citep{2020MNRAS.498.2554C, 2022ApJ...933..202X}. The observed \ion{Mg}{2} flux ratio in Makani, based on KCWI spectra, is $R \equiv F_{2796}/F_{2803} = 1.00\pm0.08$. Using the optically-thin scenario, or equivalently the clumpy scenario with optically-thin channels, from \citet{2020MNRAS.498.2554C} yields an escape fraction in the \ion{Mg}{2} 2796~\AA\ line of 25\%. Using the line fluxes from ap1.1 in the current data and the method of \citet{2022ApJ...933..202X}, we can make a second, independent measure of $f_\mathrm{esc}(2796)$. We apply the highest-metallicity model relating the intrinsic values of [\ion{O}{3}]5007/[\ion{O}{2}] and \ion{Mg}{2}2796/[\ion{O}{3}]5007 \citep{2022ApJ...933..202X} to yield $F_{2796}(\mathrm{intrinsic}) = 2.92\times10^{-16}$ erg~s$^{-1}$~cm$^{-2}$ in the optically-thin, density-bounded case. In a 1\farcs5-diameter nuclear KCWI aperture, roughly corresponding to the footprint of ap1.1, we measure $F_{2796}(\mathrm{observed}) = 0.70\times10^{-16}$ erg~s$^{-1}$~cm$^{-2}$. Then $f_\mathrm{esc}(2796) = F_{2796}(\mathrm{obs})/F_{2796}(\mathrm{int}) = 24\%$, identical to the first estimate. Comparing to observations of other \ion{Mg}{2} emitters, this value of $f_\mathrm{esc}(2796)$ suggests that $f_\mathrm{esc}(\mathrm{Lyc})$ could be a few percent or higher.

Finally, it may also be the case that the bulk of the star formation is obscured by larger columns than traced by optical light, as seen in nearby compact mergers \citep{2007ApJ...654L..49S}. Future rest-frame far-ultraviolet and mid-infrared spectra of Makani and its parent sample could distinguish among these possibilities.

The oxygen abundance of this component is 12$+$log(O/H) $\sim$ 9.1 (or $\sim$2$\times$ Solar; \citealt{2021A&A...653A.141A}), based on standard [\ion{N}{2}]/[\ion{O}{2}] and $R_{23}$ strong-line calibrations \citep{2002ApJS..142...35K}. For component 1 of ap1, [\ion{N}{2}]/[\ion{O}{2}] $= -0.12$ and $R_{23} = -0.13$. This is consistent with the high mass of the galaxy \citep{2019Natur.574..643R,2021ApJ...912...11D}. The actual abundance could be closer to Solar, as strong-line calibrations may overestimate the true gas abundance by a factor of $\sim$2 \citep{2008ApJ...681.1183K}. The weak lines [\ion{O}{3}]~4363~\AA\ and [\ion{N}{2}]~5755~\AA\ could in principle also provide a weak-line measurement based on electron temperature. However, the inferred temperatures are much too high for high-metallicity \ion{H}{2} regions, suggesting that component blending or simply an overestimate of the ap1.1 component in these weak lines is producing these high ratios.

\subsection{Comparison to shock models} \label{sec:shocks}
 
Line ratios are commonly used to distinguish between stellar and shock ionization in galactic winds \citep{1996ApJ...462..651L,2002ApJ...565L..63V,2010ApJ...711..818S}. Aside from ap1.1, the line ratios observed in Makani (Figure~\ref{fig:vo}) are inconsistent with those predicted by stellar photoionization models \citep{2001ApJ...556..121K,2017ApJ...840...44B}. An obvious place to turn instead is shock models, both due to their typical consistency with LINER-like line ratios \citep[e.g.,][among numerous examples]{1997ApJ...490..202D} and because of the extreme outflows we detect in the Makani nebula.

We first compare the position of apertures ap3, ap4, and ap5 in Figure~\ref{fig:vo} to the low-velocity range of fast shock models \citep{2008ApJS..178...20A}. We assume solar metallicity, and the benchmark model pre-shock density $n_\mathrm{H} = 1$~cm$^{-3}$ (their model series M\_n1). In the [\ion{N}{2}]/H$\alpha$ diagram, shock velocities $v_\mathrm{s}\sim200$~km~s$^{-1}$ and low magnetic field strengths $B/n^{1/2} < 0.5$~$\mu$G~cm$^{-3/2}$ can reproduce the results, whether the emission comes from the shock only or from the shock and precursor. The precursor is the gas that is about to be impacted by the shock, and which in this case is fully pre-ionized by the shock radiation. For [\ion{S}{2}]/H$\alpha$, the same low $v_s$ and low $B$ fit best in the shock-only grid, but the shock$+$precursor grid does not overlap the observed values. The [\ion{O}{1}]/H$\alpha$ grids allow for shock or shock$+$precursor solutions, though with slightly higher velocity, $v_\mathrm{s}\sim$250~km~s$^{-1}$.

The fast shock models cannot match the low observed ionization parameters from [\ion{O}{3}]/[\ion{O}{2}], even in the lower-ionization shocked region. This may point to only partial pre-ionization of the precursor, as assumed in slow shock models \citep{2010ApJ...721..505R, 2017ApJS..229...35D}, or additional post-shock physics. The slow-shock models are consistent with the observed [\ion{S}{2}]/H$\alpha$ and [\ion{O}{1}]/H$\alpha$ at the highest model velocities, $v_\mathrm{s}\sim200$~km~s$^{-1}$, but produce values of [\ion{N}{2}]/H$\alpha$ that are too high at the same velocities.

The fast shock models are mostly consistent with the observed values of [\ion{O}{2}]/H$\alpha = 1-5$ (Figure~\ref{fig:o2ha}). For shocked gas only, the model ratio is 2--4 for $v_\mathrm{s}=100-300$~km~s$^{-1}$, with a minimum around 200~km~s$^{-1}$. Including precursor, the model ratio is slightly smaller, in the 1.8--2.8 range. The observed values are much larger than those in nearby star-forming galaxies, for which the maximum observed [\ion{O}{2}]/H$\alpha$ is about 2 \citep{2006ApJ...642..775M}. It is also larger than observed in the diffuse ionized gas in extraplanar regions of nearby galaxies \citep{2017A&A...599A.141J}.

We estimate the emitted H$\beta$ fluxes, $F_\mathrm{H\beta}^\mathrm{shock}$, across the shock in the context of these models by dividing the luminosity in each aperture (Table~\ref{tab:flux}) by the projected physical area of each aperture in cm$^{-2}$ (Table~\ref{tab:aps}). This assumes the shock is plane-parallel to the line-of-sight, and so may be an underestimate if the aperture is tracing the edge of the wind. However, given the large size of the wind, projection effects are probably not significant for the inner apertures (ap1--ap5). The resulting shock fluxes $F_\mathrm{H\beta}^\mathrm{shock}$ in ap3--ap5 are approximately 10$^{-4.0}$ to 10$^{-3.7}$~erg~s$^{-1}$~cm$^{-2}$. We then compare these observed fluxes to those predicted by the the $Z = Z_\odot$ models M\_n1 and V\_n10 from \citet{2008ApJS..178...20A}. We use the shock velocity and magnetic field parameter estimated from comparing the observed line ratios to the same models, $v_\mathrm{s}=200$~km~s$^{-1}$ and $B/n^{1/2} < 0.5$~$\mu$G~cm$^{-3/2}$. (The result is not sensitive to $B$.) Finally, interpolating between the model grids, the extrapolated pre-shock density from fast models is then $n_\mathrm{H} = 2-3$~cm$^{-3}$, since the flux scales linearly with $n_\mathrm{H}$ \citep{1995ApJ...455..468D}. At larger radius, the H$\beta$ fluxes decrease to the range $10^{-4.3}$ to $10^{-5.7}$~erg~s$^{-1}$~cm$^{-1}$, which suggests an increase in projection effects that mean these are lower limits; or a lower velocity and/or density at these radii. Lower velocity would be consistent, for instance, with the lower line ratios in ap6 and ap8 compared to ap3--ap5, but it seems probable that the gas density would also decrease with radius.

At $r=$ 0--5~kpc, the high-velocity outflow component ap1.2 shows elevated [\ion{O}{3}]/H$\beta$, [\ion{N}{2}]/H$\alpha$ and [\ion{O}{3}]/[\ion{O}{2}]; and lower [\ion{O}{1}]/H$\alpha$ and [\ion{S}{2}]/H$\alpha$. The H$\beta$ flux is also higher at 10$^{-2.65}$~erg~s$^{-1}$~cm$^{-1}$. The line ratios place this broad component in the AGN regions of the excitation diagrams. However, $n_\mathrm{H}\sim10$~cm$^{-3}$ shock$+$precursor models from \citet{2008ApJS..178...20A} (their models V\_n10) are entirely consistent with the observed line ratios if $v_\mathrm{s}$ is in the range 300--400~km~s$^{-1}$ and $B/n^{1/2} < 0.5$~$\mu$G~cm$^{-3/2}$. These models also reproduce the observed ratios [\ion{Ne}{5}]~3426/H$\beta=(-1.33\pm0.20)$~dex and [\ion{Ne}{5}]~3426/[\ion{Ne}{3}]~3869 $=(-0.45\pm0.22$)~dex. The higher ionization state is due to the harder radiation field produced by the higher-velocity shocks. This harder radiation also produces high temperatures in the precursor and post-shock regions. The observed [\ion{O}{3}]~4363/5007 ratio in ap1.2 is -1.1~dex, consistent with these high temperature ($T\ga 2\times10^4$~K; Section \ref{sec:results}). Such ratios are commonly observed in LINERs \citep{2001ApJ...549..155N, 2018ApJ...864...90M}. However, the models cannot perfectly reproduce the combined [\ion{O}{3}]~4363/5007 and [\ion{O}{3}]/H$\beta$ ratios, an ongoing problem \citep{1995ApJ...455..468D,2008ApJS..178...20A}.

In summary, the fast shock models of \citet{2008ApJS..178...20A} are very consistent with the observed strong line ratios and fluxes throughout the nebula. In ap1.2 (the outflow at radii 0--5~kpc), high shock velocities (300--400~km~s$^{-1}$) and an ionized precursor are needed. These shock velocities align with the observed $\langle \sigma \rangle_\mathrm{II} = 400$~km~s$^{-1}$ \citep{2019Natur.574..643R}. Farther from the host, the required shock velocities are lower, 200~km~s$^{-1}$, but these still align with the observed linewidths $\langle \sigma \rangle_\mathrm{I} = 200$~km~s$^{-1}$. The precursor emission contributes fractionally less at these velocities. In the next section, we apply the shock models to constrain the wind density.

\subsection{Shock structure and wind density} \label{sec:winddensity}

We can combine the observed line flux ratios [\ion{S}{2}]6716/6731 and [\ion{O}{2}]3729/3726 with the \citet{2008ApJS..178...20A} shock models to constrain the ambient, pre-shock and post-shock densities in the wind and extended nebula. The high-velocity Episode II wind seen in ap1--ap3 has a high electron density $n_e = 3000$~cm$^{-3}$ (1$\sigma$ range 1200--10,000~km~s$^{-1}$) from the [\ion{S}{2}] ratio of ap1.2. This density is also consistent with the [\ion{O}{2}] ratio. The shock model that best fits the observed line ratios and shock fluxes in ap1.2--ap3 has a pre-shock density of $n_\mathrm{H}\sim10$~cm$^{-1}$, as we discuss above in Section~\ref{sec:shocks}. However, compression in the shock raises the density in the post-shock region up to a factor of $n_\mathrm{post}/n_\mathrm{pre} =  (8\pi\mu m_\mathrm{H})^{1/2} v_\mathrm{shock}/(B/n^{1/2})$ \citep{1996ApJS..102..161D}. For the estimated velocities of 300--400~km~s$^{-1}$ and low magnetic field parameter $0.5~\mu$G~cm$^{-3/2}$, the maximum value of log($n_\mathrm{post}/n_\mathrm{pre})$ is 2.75. This translates to a post-shock density of at most $n_\mathrm{post} \sim 6000$~cm$^{-3}$ for $n_\mathrm{pre} = 10$~cm$^{-3}$. (The compression can also be seen in Figures 3--6 of \citet{2008ApJS..178...20A}.)

The correspondence between the observed density and predicted post-shock density suggests that the observed [\ion{S}{2}] and [\ion{O}{2}] emission is arising primarily in the compressed post-shock region. The line ratios, however, appear to require an ionized precursor. That the post-shock region seems to dominate the flux of these lines could point to additional physics occurring behind the shock front, perhaps due to the accumulation of swept-up interstellar gas or to the driving mechanism of the wind.

At larger radii, in the extended, Episode I nebula, the line ratios and fluxes point to much lower ionized gas densities $n_e \la 10$~cm$^{-3}$. The shock models are consistent with low pre-shock densities of $n_\mathrm{H} \la 2-3$~cm$^{-3}$, with this value possibly decreasing with increasing radius. The corresponding maximum post-shock density is $n_\mathrm{H} \la 600$~cm$^{-3}$. In contrast to the high-velocity, high-density Episode~II gas, this implies that the [\ion{S}{2}] ratio is primarily tracing the lower-density precursor gas. The line ratios are, for the most part, consistent with either shock-only or shock$+$precursor models.

The 200 and 400~km~s$^{-1}$ models with $n_\mathrm{H}=1-10$~cm$^{-3}$ (models M\_n1 and V\_n10) predict that just a little over half of the H$\beta$ luminosity is produced in the post-shock gas, while the other fraction arises in the precursor, with little dependence on magnetic parameter $B/n^{1/2}$ \citep{2008ApJS..178...20A}. Thus, in the calculations that follow, we assume complete ionization in each shock region and divide the flux evenly between these low- and high-density regions. Using the estimated pre-shock densities and calculated maximum compression as a guideline, we assume $n_e = 10$ and 1000~cm$^{-3}$ pre- and post-shock for Episode I, and $n_e = 1$ and 100~cm$^{-3}$ for Episode II. The post-shock densities could be even higher, but higher values do not significantly change the results, as the lower-density pre-shock gas dominates the mass.

\begin{deluxetable*}{cccccccccc}
\tablecaption{Ionized Wind Properties\label{tab:wind}}
  \tablewidth{0pt}
  \tablehead{\colhead{Ep} & \colhead{Method} & \colhead{$L_\mathrm{H\alpha}^\mathrm{wind}$} & \colhead{M} &  \colhead{$\langle v_\mathrm{rad} \rangle$} & \colhead{dM/dt} & \colhead{p} & \colhead{dp/dt} & \colhead{E} & \colhead{dE/dt} \\
  \colhead{~} & \colhead{~} & \colhead{erg~s$^{-1}$} & \colhead{M$_\odot$} & \colhead{km~s$^{-1}$} & \colhead{M$_\odot$~yr$^{-1}$} & \colhead{dyn~s} & \colhead{dyn} & \colhead{erg} & \colhead{erg~s$^{-1}$} \\
    \colhead{(1)} & \colhead{(2)} & \colhead{(3)} & \colhead{(4)} & \colhead{(5)} & \colhead{(6)} & \colhead{(7)} & \colhead{(8)} & \colhead{(9)} & \colhead{(10)}}
  \startdata
I   &  1a &  (2.25$^{+2.25}_{-1.13}$)E+42 &  5.09E+09 &    98 &    9.8 &  7.59E+49 &  6.92E+33 &  6.90E+57 &  4.23E+41 \\
I   &  1b &  2.25E+42 &  5.09E+09 &   622 &   64.8 &  5.04E+50 &  2.45E+35 &  2.16E+58 &  1.58E+43 \\
I   &   2 &  2.25E+42 &  5.09E+09 &    97 &   12.7 &   \nodata &   \nodata &   \nodata &   \nodata \\
\hline
II  &  1a &  (5.75$^{+3.52}_{-1.76}$)E+42 &  1.30E+09 &   157 &    9.2 &  2.70E+49 &  1.55E+34 &  7.27E+57 &  2.55E+42 \\
II  &  1b &  5.75E+42 &  1.30E+09 &  2279 &  151.4 &  4.42E+50 &  2.55E+36 &  6.59E+58 &  8.77E+44 \\
II  &   2 &  5.75E+42 &  1.30E+09 &  2091 &  185.8 &   \nodata &   \nodata &   \nodata &   \nodata \\
\hline
  \enddata
  \tablecomments{Properties of the ionized wind for each episode are listed for three different methods of calculation, as described in Section~\ref{sec:mpe}. Methods 1a and 1b make use of the outflow dynamical timescale to calculate the mass, momentum, and energy outflow rates, while method 2 is based on the stellar population ages for episodes I and II. For Episode I, we assume a wind radius $R_\mathrm{I} = 40$~kpc, electron density $n_e = 1-100$~cm$^{-3}$, and time when the wind left the starburst (for Method 2) equal to the stellar population age $t_{*,\mathrm{I}} = 400$~Myr. For Episode II, we assume $R_\mathrm{II} = 15$~kpc, $n_e = 10-1000$~cm$^{-3}$, and $t_{*,\mathrm{II}} = 7$~Myr. We also remove 40\%\ of the flux at $r < 5.5$~kpc to conservatively account for star formation in the inner nebula (Section~\ref{sec:starburst}), which is equivalent to removing $1.4\times10^{42}$~erg~s$^{-1}$ from Episode II. Our preferred methods are 1a and 2 for Episode I; and 1b and 2 for Episode II. The 1$\sigma$ errors propagated from our bootstrapping method (Section~\ref{sec:results}) are included for $L_\mathrm{H\alpha}$, and similar fractional errors apply to other quantities calculated from $L_\mathrm{H\alpha}$. Systematic errors due to our choice of density model, etc., are not included.}
\end{deluxetable*}
\subsection{Mass, Momentum, and Energy} \label{sec:mpe}

As the KCWI observations did not cover strong H recombination lines,  we were previously unable to measure the mass of the ionized Makani nebula. Here we use our bootstrapped estimate of the spatially-resolved H$\alpha$ luminosity of the wind (Section~\ref{sec:results}) to estimate the mass and dynamics of both the Episode I and II winds.

We start with the Voronoi-binned, KCWI [\ion{O}{2}] map and spatially delineate the Episode I and II winds using a velocity cut at $v_{98\%} = -700$~km~s$^{-1}$ \citep{2019Natur.574..643R}. To do this, we characterize multi-component line fits in the KCWI data with the cumulative velocity distribution. We calculate velocities that enclose a particular percentage p of flux integrated from the red side of the line, $v_\mathrm{p\%}$. (E.g., $v_{98\%}$ is the velocity at which 98\%\ of the line flux is redshifted from this velocity, and so characterizes the maximum blueshift observed.) We compute the H$\alpha$ flux and luminosity in each bin, applying our model for ([\ion{O}{2}], observed)/(H$\alpha$, intrinsic) vs. radius and using the mean radius in each Voronoi bin. From the inner $r<5.5$~kpc, we also conservatively remove 40\%\ of the flux in each bin for the contribution from star-forming gas (Section~\ref{sec:starburst}).

As discussed in Section~\ref{sec:winddensity}, we use the observed line ratios and shock models to infer how much of the observed luminosity arises from regions of different density. Using these luminosities and densities, we then calculate the ionized gas mass in each bin.  This ionized density model differs from many estimates in the literature, which typically use only the density computed from line ratios. However, the shock models provide further information on the density structure that could yield a more accurate estimate. Given that $M^\mathrm{HII}\sim n_e^{-1}$, the masses are largely dependent on the lower, pre-shock density in our model. This lower density is mostly consistent with the measured densities (Section~\ref{sec:results}). However, the pre-shock density in the Episode II shock model is lower than the density measured in the densest part of the wind, in ap1.2, by a factor of $\sim$100. Assuming a constant density wind with the higher density measured from [\ion{S}{2}] and [\ion{O}{2}] in ap1.2 would result in a mass outflow rate that is lower by a similar factor.

The resulting H$\alpha$ luminosities and gas masses are listed in Table~\ref{tab:wind}. The Episode I gas is fainter but contains more mass because of our density model. Together, the total ionized gas mass of $6.4^{+5.2}_{-2.6}\times10^{9}$~M$_\odot$ is comparable to the integrated CO mass of $1\times10^{10}$~M$_\odot$ \citep{2019Natur.574..643R}. The detected CO is much less extended than the ionized gas (reaching a radius of only 20 kpc; Figure~\ref{fig:neutral}). The Episode II ionized gas mass is also comparable to the CO mass of $(2.4\pm0.6)\times10^{9}$~M$_\odot$ in a similar velocity range \citep{2019Natur.574..643R}.

To compute outflow rates (Table~\ref{tab:wind}), we use two methods. First, we assume a single-radius wind for each episode, using $R_{I}=40$~kpc and $R_\mathrm{II}=15$~kpc (Method 1). We use both the central velocity $v_{50\%}$ (Method 1a) and maximum velocity $v_{98\%}$ (Method 1b) to compute the wind properties. We deproject the velocity (either $v_{50\%}$ or $v_{98\%}$) in each bin to compute the 3D $v_\mathrm{rad}$, using the projected galactocentric distance of the bin $R_\mathrm{obs}$ and assuming that it lies on the wind surface at the assumed 3D radius $R_\mathrm{wind}$: $v_\mathrm{rad} = v_{obs} / \mathrm{cos}[\mathrm{sin}^{-1}(R_\mathrm{obs}/R_\mathrm{wind})]$. We then compute the ballistic flow time for each bin, $t_\mathrm{flow} = v_\mathrm{rad} / R_\mathrm{wind}$ and divide the mass in that bin by $t_\mathrm{flow}$. We follow a similar procedure for momentum and energy, adding in velocity dispersion for energy (following \citealt{2005ApJS..160..115R}).

As an alternative method we assume the stellar population ages $t_*$ for Episodes I and II, 400 and 7~Myr, represent the times when all of the gas in each episode left the starburst (Method 2). We then divide the ionized mass by these times for each episode to compute $dM/dt=M/t_*$. This method is effectively a time-averaged outflow rate, where the time average is over the lifetime of the outflow. We use it as a sanity check on Method I.

For Episode I, Method 1a (using $v_{50\%}$) yields a result consistent with Method 2, as the mean deprojected radial velocity $\langle v_\mathrm{rad} \rangle = 90$~km/s is close to the velocity inferred from $R_\mathrm{I}/t_\mathrm{*,I} = 40~\mathrm{kpc}/400~\mathrm{Myr} = 97$~km/s.  In other words, most of the gas is at a velocity consistent with the outer radius of the nebula and the ballistic flow velocity, if this gas left the starburst 400~Myr ago. The gas that began at higher velocity has either slowed or since escaped to much larger radius.  The remaining detected amounts of high-velocity gas do not contribute significantly to the mass, or mass outflow rate, of the outflow. Using $v_{98\%}$ to trace the radial velocity of the bulk of the gas thus overestimates the flow rate.

For Episode II, Method 1b (using $v_{98\%}$) is consistent with Method 2, due again to comparable $\langle v_\mathrm{rad} \rangle = 2279$~km/s and $R_\mathrm{II}/t_{*,\mathrm{II}} = 15~\mathrm{kpc}/7~\mathrm{Myr} = 2090$~km/s. Method 1a, in contrast, yields a flow rate 15--20$\times$ lower. In this case, while the fastest gas has flowed to the largest observed radius, the lower velocity gas has not had time to do so. Thus, the assumption of a single radius for the inner wind is probably incorrect, and a flow rate computed from the flow time in each spaxel, $t_*$ (i.e., Method 2), is more correct. This intepretation implies that the inferred radius, $R_\mathrm{II} = v_\mathrm{rad} / t_{*,\mathrm{II}}$, is velocity-dependent. Lower-velocity spaxels will have smaller radius, and our single-radius deprojection will underestimate the radial velocity. For both of these reasons, Method 1a produces a significant underestimate for the Episode~II wind. 

We conclude that Methods 1a and 2 are closer to the correct flow rates for Episode I, while Methods 1b and 2 are closer to the correct flow rates for Episode II. The inferred mass outflow rates are thus in the range 10--13~M$_\odot$~yr$^{-1}$ for Episode I and 151--186~M$_\odot$~yr$^{-1}$ for Episode II. Again, $dM/dt_\mathrm{II}$ for the ionized gas is comparable to the CO value, 245~M$_\odot$~yr$^{-1}$. In most nearby compact starbursts, $dM/dt$ in the ionized gas is much smaller than in the molecular gas, though with significant scatter \citep[e.g.,][]{2013ApJ...768...75R,2021MNRAS.505.5753F}. However, the canonical density model in these calculations is inferred from optical line ratios; we combine the line ratio densities with the shock model density structure, which raises the inferred masses. In Perrotta et al. (2022, in prep.), we will show that measurements of $dM/dt$ from \ion{Mg}{2} and \ion{Fe}{2} absorption lines in other compact starbursts from the \citet{2007ApJ...663L..77T} sample are consistent with mass outflow rates of order $10^{2-3}$ M$_\odot$~yr$^{-1}$.

Note that we choose the single wind radius for each episode based on two factors. The first is the observed wind morphology \citep{2019Natur.574..643R}. We refine the radius to approximately match the average deprojected velocity $\langle v_\mathrm{rad} \rangle$ and the velocity $R / t_*$ for the two methods (1a or 1b and 2) that produce the best agreement in $dM/dt$ for each episode. The resulting $dM/dt$ is not sensitive to a 30\%\ change in $R_\mathrm{wind}$.

\subsection{Powering the Nebula and Driving the Wind}

We have shown that the line emission from the wind is most consistent with shock models (Section~\ref{sec:shocks}). We nonetheless consider if the radiative energy from the starburst is capable of powering the H$\alpha$ luminosity of the nebula.

We first assume the current SFR as given by IR/radio tracers, rather than H$\alpha$. A $\la$6.5~Myr instantaneous burst or a $\ga$7~Myr continuous burst with star formation rate of 224--300 $M_\odot$~yr$^{-1}$ produces $Q(\mathrm{H}^0) = (1.9-2.6)\times10^{55}$ ionizing photons s$^{-1}$ \citep{1999ApJS..123....3L}. The ratio of H$\alpha$ photons to ionizing photons is $\alpha_\mathrm{H\alpha}^\mathrm{eff}/\alpha$, or the ratio of the effective to total recombination coefficients. For Case B at $T = 10^4$~K and $n_e = 100$~cm$^{-2}$, $\alpha_\mathrm{H\alpha}^\mathrm{eff}/\alpha_\mathrm{B} = \alpha_\mathrm{H\beta}^\mathrm{eff}/\alpha_\mathrm{B}\times(j_\lambda\lambda)_\mathrm{H\alpha}/(j_\lambda\lambda)_\mathrm{H\beta} = 0.452$, where $j_\lambda$ are the line emissivities and we use the emissivity and recombination coefficient tabulations of \citet{1987MNRAS.224..801H}. This yields an H$\alpha$ luminosity predicted from the radio/IR SFR of $L_\mathrm{H\alpha}^\mathrm{SFR}=(2.6-3.5)\times10^{43}$~erg~s$^{-1}$, which is $3-4$$\times$ larger than the observed luminosity of the nebula, $L_\mathrm{H\alpha}^\mathrm{tot}(\mathrm{H}\alpha)=9.4\times10^{42}$ erg~s$^{-1}$ (Section \ref{sec:results}).

Photoionizing the entire nebula would require a significant fraction of the Lyman continuum to leak outside of the inner starburst region. We can infer from the predicted H$\alpha$ luminosity $L_\mathrm{H\alpha}^\mathrm{SFR}$ and the H$\alpha$ emission attributed to the starburst, $L_\mathrm{H\alpha}^\mathrm{starburst}= 1.8\times10^{42}$~erg~s$^{-1}$ (Section~\ref{sec:starburst}), that a fraction $L_\mathrm{H\alpha}^\mathrm{starburst}/L_\mathrm{H\alpha}^\mathrm{SFR}\sim0.05-0.07$ of the LyC is required to ionize the gas consistent with stellar photoionization (ap1.1).  To energize the H$\alpha$ emisison in the wind ($L_\mathrm{H\alpha}^\mathrm{wind} = 8.0\times10^{42}$~erg~s$^{-1}$; Table~\ref{tab:wind}) would require an additional 23--31\%\ of the LyC emitted by the starburst. The total escape fractions for the inner starburst and entire nebula, respectively, would then be of order 90\%\ and 65\%, which are uncomfortably high for a galaxy of this mass and ionization parameter (Izotov et al. 2022).
Thus, if the nebula were photoionized by the starburst rather than shocks, significant absorption of LyC by dust or a very recent SFR that is much lower would be necessary (Section \ref{sec:starburst}).

The shock models also make energetic predictions about the total flux radiated in the shocks to which we can compare the data. The mechanical energy flux that powers the shock is completely radiated by the cooling gas \citep{1996ApJS..102..161D}. Assuming H$\alpha$/H$\beta=2.86$, the ratio of the total radiated luminosity $L_\mathrm{rad}$ to the observed H$\alpha$ luminosity is given by \citep{1995ApJ...455..468D}

\begin{equation}
\frac{L_\mathrm{shock}}{L_{\mathrm{H}\alpha}} = 107~v_\mathrm{s,100}^{0.59}~(1+1.32~ v_\mathrm{s,100}^{-0.13})^{-1}
\end{equation}

\noindent where $v_{\mathrm{s},100}$ is the shock velocity in units of 100~km~s$^{-1}$.

Here we compare the energy produced in the shocks to the mechanical energy observed in the warm, ionized and molecular phases of the wind episodes. We also look at how the mass, momentum and energy in the wind could arise from the starburst itself.

\subsubsection{Episode II ionized wind}

The best-fitting model shock velocity is $v_\mathrm{s}=400$~km~s$^{-1}$, which yields $L_\mathrm{shock,II}=115L_{\mathrm{H}\alpha,\mathrm{II}}^\mathrm{wind}=6.6\times10^{44}$~erg~s$^{-1}$. If we assume, based on their similar mass and dynamics, that the ionized and molecular components of the fast wind contribute similarly to the energy flow rate $dE/dt$, then $dE/dt_\mathrm{II}^\mathrm{HII+H_2} = 1.8\times10^{45}~\mathrm{erg~s}^{-1}=2.7 L_\mathrm{shock,II}$. In other words, the inferred mechanical energy in the ionized and molecular components of the wind is capable of powering the total radiated luminosity in the shock models.

The mechanical energy produced by the starburst is probably less than both of these numbers. Assuming a continuous starburst of age $t_\mathrm{*,II}$, Solar metallicity, and a Salpeter IMF with stellar masses $M = 0.1-100$~M$_\odot$, Starburst99 predicts a mechanical luminosity in the hot ejecta of $dE/dt_* = 2.4\times10^{43}$~erg~s$^{-1}$ \citep{1999ApJS..123....3L}. Because of the relatively young age of this starburst, some SNe have not yet released their energy. So older, continuous bursts would contain a greater fraction of the asymptotic value of $dE/dt_*$, which is $\sim$$8\times10^{43}$~erg~s$^{-1}$ for $\mathrm{SFR} = 300$~M$_\odot$~yr$^{-1}$. Even if we assume that 100\%\ of this mechanical energy is thermalized into hot ejecta that then drives the outflow, the range of possible $dE/dt_*$ is $\la$4\%\ of $dE/dt_\mathrm{II}^\mathrm{HII+H_2}$.

An alternative to being driven by the energy in the hot ejecta from the central starburst is that the Makani wind is driven by the mechanical and radiative momentum of the burst. Compact starbursts with strong winds, like Makani, are prime candidates for outflows driven by the radiation pressure from Eddington-limited star formation \citep{2005ApJ...618..569M,2005ApJ...630..167T,2012ApJ...755L..26D}. The momentum in the stellar winds and supernova ejecta from the starburst is $dp/dt_* = \sqrt{2~dM/dt_*~dE/dt_*}$ \citep{2020A&ARv..28....2V}. The input mass loss rate from the hot ejecta is $dM/dt_* \sim 12$~M$_\odot$~yr$^{-1}$ \citep{1999ApJS..123....3L}, so that $dp/dt_* = 1.9\times10^{35}$~dyn. However, observations of the hot wind eject in M82 suggest that $dM/dt_* = (0.1-0.3) \mathrm{SFR}$ \citep{2009ApJ...697.2030S}, depending on the mass-loading of the hot wind by cold clouds within the starburst region. This would imply $dM/dt_* \sim 30-90$~M$_\odot$~yr$^{-1}$, and in turn $dp/dt_* = (3-5)\times10^{35}$~dyn. For the calculations below, we take the midpoint of this range, $4\times10^{35}$~dyn.

The input from radiation pressure is $dp/dt_\mathrm{rad} = (1+\tau_{\mathrm{eff,IR}})L_\mathrm{bol}/c$, assuming an optically-thick wind, where $\tau_\mathrm{eff,IR}$ is the effective far-infrared optical depth and accounts for multiple photon scatterings \citep{2015MNRAS.449..147T,2020A&ARv..28....2V}. For Makani, $dp/dt_\mathrm{rad}=(1+\tau_{\mathrm{eff,IR}})3.8\times10^{35}$~dyn, based on  $L_\mathrm{bol} \sim 2L_\mathrm{IR} = (1.16\pm0.18)\times10^{46}$~erg~s$^{-1}$ \citep{2020ApJ...901..138P}.

We find $dp/dt_\mathrm{II}^\mathrm{HII+H_2} = 7\times(dp/dt_* + dp/dt_\mathrm{rad})$ for $\tau_\mathrm{eff,IR} = 0$. A similar momentum ``boost'' is commonly observed in molecular AGN outflows \citep{2011ApJ...733L..16S,2014A&A...562A..21C,2017ApJ...836...11G}; in that context, the boost is defined as  $(dp/dt)/(L/c)$ . Boosts $(dp/dt)/(dp/dt_*) = 1-10$ are common in the warm, ionized outflows of 10--100~M$\odot$~yr$^{-1}$ starbursts \citep{2015ApJ...809..147H}. As we discuss above, the energy delivered by the starburst hot ejecta may be insufficient to drive the wind. The bulk momentum $dp/dt_*$ of the hot wind may be transferred to the cold clouds, but additional momentum can come from work done by the expanding hot bubble during its energy-conserving phase of expansion \citep{2018MNRAS.481.1873L, 2020A&ARv..28....2V}. The maximum momentum of a wind powered by a single-phase, radiatively cooling hot wind is also predicted to be several times higher than the estimate of $dp/dt_*$ based on M82, at 1.2$\times10^{36}$~dyne \citep{2021MNRAS.504.3412L}. When combined with the momentum from radiation pressure, this is only a factor of two below the observed momentum flow rate of the wind. Finally, $\tau_\mathrm{eff,IR} \gg 1$ (i.e., many photon scatterings) could in principle contribute significantly to driving the wind (\citealt{2017ApJ...839...54Z}; though see also \citealt{2022MNRAS.517.1313M}). This would be consistent with the dusty, molecular$+$neutral phase observed in the Episode~II flow (Section~\ref{sec:dustywind}).

The gravitational potential of the galaxy will work to decelerate this flow. At a radius of 10~kpc, $F_\mathrm{grav} \sim 1\times10^{35}$~dyne for the $4\times10^{9}$~M$_\odot$ of outflowing molecular and ionized gas we observe. Thus, gravity is not significantly decelerating the fast, Episode~II wind. Since $F_\mathrm{grav} \propto M_\mathrm{wind}/r^2$, at smaller radius the wind may have experienced significant deceleration, but it also may have entrained much less material at earlier times.

Momentum is certainly flowing from the hot to the cool wind. A substantial fraction of the cool, outflowing phase thus arises from clouds that are accelerated by the hot phase. The cool clouds may also acquire mass directly from the hot wind and CGM surrounding the galaxy. The hot wind can transfer mass to the cold phase via radiative cooling \citep{2016MNRAS.455.1830T} and/or turbulent mixing at cloud boundaries \citep{2018MNRAS.480L.111G, 2020MNRAS.492.1970G, 2020ApJ...894L..24F, 2020ApJ...895...43S}.

The mass outflow rate in the Episode II wind implies significant entrainment of cool gas, in that the ratio of mass flowing out in the cool, ionized and molecular wind is many times the mass predicted to be in the hot wind originating in the starburst. Using the values of $dM/dt_*$ from \citet{2009ApJ...697.2030S} that assume some mass-loading of the hot ejecta from stellar winds and SNe within the starburst injection zone (see above), $(dM/dt_\mathrm{II}^\mathrm{HII+H_2})/(dM/dt_*)=5-15$. The mass outflow rate is also larger than the star formation rate: $dM/dt_\mathrm{II}^\mathrm{HII+H_2} \sim 1.4$~SFR for SFR $=300$~M$_\odot$~yr$^{-1}$. This value of  $(dM/dt_\mathrm{II})/\mathrm{SFR}$ is consistent with other measurements of star-forming galaxies of similar mass \citep{2015ApJ...809..147H}, which range from 0.3 to 3. Though the $(dM/dt_\mathrm{II})/(dM/dt_*)$ ratio is conceptually closer to representing the mass loading of the hot wind with cool clouds, the ratio $(dM/dt_\mathrm{II})/\mathrm{SFR}$ is also called the ``mass loading factor.'' The degree of cold cloud entrainment in analytic two-phase models then points to large initial cloud sizes $M_\mathrm{cl} \ga 10^{5-6}$~M$_\odot$ in the Makani wind, such that the cooling time is short compared to the mixing time \citep{2022ApJ...924...82F}. It also implies that the large observed cool cloud ``mass loading'' of $(dM/dt_\mathrm{II})/SFR \sim 1.4$ is not, in fact, too large to cause the wind to stall, as suggested by some energy-driven models \citep{2016MNRAS.455.1830T, 2022ApJ...924...82F}. This is probably consistent with the need for significant radiation-pressure driving of the cool wind.

This mixing of mass and energy between the cool and hot phases could also be an alternative power source for the ionized nebular emission. In the single-phase analytic model of \citet{2016MNRAS.455.1830T}, the hot wind cools radiatively as it expands into the galaxy, and then again as it shocks with the surrounding CGM. By contrast, in the explicitly multiphase analytic model of  \citet{2022ApJ...924...82F}, the cool phase exists even at the base of the wind. The multiphase wind carries millions of $n\la10$~cm$^{-3}$ clouds that are area-filling (though not volume-filling). The cooling gas in the cloud boundary layer between the hot and cool phases, with $T = 0.2-1\times10^5$~K, then produces line emission in the wind.

Increasing $n_e$ in the ionized gas, so that all of the H$\alpha$ emission arises in the compressed, post-shock region (Section~\ref{sec:winddensity}), would lower the mass, momenta, and energy in the ionized gas by a factor $\sim$50. This would reduce the observed momentum boost and mass loading factor. However, the fact would remain that the mass, momentum, and energy in the molecular gas are still at the level shown in Table~\ref{tab:wind} for Episode II, so that the total flow rates would only decrease by a factor of order 2. Furthermore, the large flow rates we observe are consistent with those measured in the neutral and ionized phase of similar galaxies via other probes (Perrotta et al. 2022, in prep.).

\subsubsection{Episode I ionized wind}

The Episode I ionized wind, with $v_\mathrm{s}=200$~km~s$^{-1}$, has an estimated ionized mass $M_\mathrm{I}\sim4M_\mathrm{II}$. This mass may be distributed over a much larger volume, however: [$(R_\mathrm{I}/R_\mathrm{II})^3-1]\sim15$. The outflow rate in Episode I is also more than 10$\times$ smaller. As discussed above, high-velocity gas from Episode~I may have carried significant amounts of mass beyond $r_\mathrm{wind}=50$~kpc, further into the CGM. Alternatively, the wind may simply have slowed down in the extended halo/CGM of the galaxy, reducing $dM/dt$ while still entraining more gas at larger radius \citep{2018MNRAS.481.1873L}. Consistent with this possibility is the observation that the outflow velocity in the parent sample of Makani decreases with increasing light-weighted age (Davis et al. 2022, submitted). In Episode II, some or all of the cool wind arose from the host galaxy, originating in cool clouds or condensed from the hot wind. However, much more of the mass entrainment at larger radii may be due to radiative cooling and/or turbulent mixing from the hot wind and/or CGM \citep{2016MNRAS.455.1830T,2018MNRAS.480L.111G, 2020ApJ...894L..24F}.

The radiated energy in the Episode I shocks is $L_\mathrm{shock,I}=73L_{\mathrm{H}\alpha,\mathrm{I}}=1.6\times10^{44}$~erg~s$^{-1}$. For Episode I, we include only the ionized gas, since the CO gas is contained in the Episode II footprint. Thus, $dE/dt_\mathrm{I}^\mathrm{HII} = 0.003 L_\mathrm{shock,I}$, a significant discrepancy. We conclude that much of the energy in the shocks beyond $r\sim20$~kpc must come from the mechanical energy of a different phase of the outflow. This could be the hot phase traced by higher ionization states like \ion{O}{6}, as observed in the CGM in absorption at large impact parameters \citep[e.g.,][]{2011Sci...334..948T}. Shocks may in fact arise naturally in the interaction between a hot wind and surrounding CGM \citep{2016MNRAS.455.1830T}.

\subsection{Episode II dusty wind} \label{sec:dustywind}

Observations of molecular gas and \ion{Mg}{2}~2796,~2803~\AA\ emission to radii 20--25~kpc point to the presence of significant \ion{H}{1} and H$_2$ in the inner, Episode~II wind.  The presence of these two tracers also indicates significant amounts of dust out to similar radii. However, these dusty gas phases appear only at the inner edge of the Episode~I wind \citep{2019Natur.574..643R}.

Two new lines of evidence confirm this picture. First, \ion{Na}{1}~D is detected in two apertures out to $r = 10-15$~kpc (Figure~\ref{fig:neutral}). The presence of \ion{Na}{1}~D has long been closely associated with dust extinction \citep[see, e.g.,][and references therein]{2021MNRAS.503.4748R}, due to dust being required to keep Na in the neutral phase. Second, we detect substantial gas extinction, $E(B-V)\sim0.5$, out to $r=20-25$~kpc (Figure~\ref{fig:o2ha}).

Such dusty, neutral outflows on $\sim$10~kpc scales are commonly observed in compact starbursts in nearby mergers \citep{2013ApJ...768...75R}. Frequently, high spatial resolution observations show filamentary dust structures associated with these outflows. Future high-resolution observations of Makani with the {\it James Webb Space Telescope} may resolve these dust structures, as well as more directly determine the radial extent of dust in the wind.

\section{Conclusion} \label{sec:conclude}

The giant Makani galactic wind is driving ionized, neutral, and molecular gas and dust out of a compact starburst into the CGM. In the preceding sections we have presented a spatially-resolved analysis of the physical state and mass, momentum, and energy in the ionized phase of the wind using rest-frame optical spectra. These ionization and dynamical measurements are key to unlocking the driving force of the Makani wind, as well as its impact on and interaction with the surrounding CGM.

The fast, inner wind extending to $R_\mathrm{II}=20$~kpc---Episode II---is powered by a young starburst of age 7~Myr \citep{2019Natur.574..643R} and star formation rate 224--300~M$_\odot$~yr$^{-1}$ \citep{2020ApJ...901..138P}. Using line ratios and the luminosity of recombination lines, we conclude that the ionized gas in the inner wind is energized by fast, $v_\mathrm{s}=400$~km~s$^{-1}$ shocks moving through a low-density $n_\mathrm{H} \sim 10$~cm$^{-3}$ medium \citep{2008ApJS..178...20A}, consistent with earlier results \citep{2021ApJ...923..275P}. The velocity dispersion of the gas is equal to this model shock velocity: $\langle \sigma \rangle_\mathrm{II} = 400$~km~s$^{-1}$. The shock compresses this gas to high density, yielding $n_e \ga 10^3$~cm$^{-3}$ in the post-shock region. The hard radiation field of the shock produces high-ionization line emission commonly seen  in AGN ([\ion{Ne}{5}]~3426~\AA), as well as high gas temperatures $T>2\times10^4$~K, as traced by [\ion{O}{3}]~4363~\AA\ and [\ion{N}{2}]~5755~\AA.

Molecular and neutral gas was previously detected throughout the Episode~II wind \citep{2019Natur.574..643R}. We find new evidence for neutral, dusty gas in the wind in the form of  \ion{Na}{1}~D absorption in the host galaxy and emission out to 10--15~kpc; and extinction in the ionized gas, as traced by the Balmer decrement, to 20--25~kpc.

A much older and larger wind, extending from $R_\mathrm{I}=20-50$~kpc---Episode I---was powered by a star formation event 400~Myr ago \citep{2019Natur.574..643R}. Presently, only ionized gas is detected in the Episode~I wind, and it is energized by 200~km~s$^{-1}$ shocks in $n_\mathrm{H} \la 2-3$~cm$^{-3}$ gas. Again, this is consistent with the observed velocity dispersion of this outflow, $\langle \sigma \rangle_\mathrm{I} = 200$~km~s$^{-1}$ \citep{2019Natur.574..643R}.

Using measurements of [\ion{O}{2}]~3726,~3729~\AA\ and H$\alpha$ throughout the wind, we model the radial dependence of their ratio. We apply this model to the fully spatially-resolved [\ion{O}{2}] map of the nebula to compute $L_\mathrm{H\alpha}$ across the full nebula. We then combine the shock structure predicted by models \citep{1996ApJS..102..161D,2008ApJS..178...20A} with the density from line ratios to measure the mass in each wind episode. Finally, we employ three different methods for computing $dM/dt$, $p$, $dp/dt$, $E$, and $dE/dt$ in the ionized wind from the 2D mass and velocity maps of the nebula.

Our preferred model of the Episode~II nebula is a massive, powerful, ionized wind that is comparable to that of the molecular phase, with $M_\mathrm{II}^\mathrm{HII}\sim M_\mathrm{II}^\mathrm{H_2} = (1-2)\times10^9~\mathrm{M}_\odot$ and $dM/dt_\mathrm{II}^\mathrm{HII}\sim dM/dt_\mathrm{II}^{\mathrm{H}_2} = 170-250~\mathrm{M}_\odot$~yr$^{-1}$. Together, the ionized and molecular phases are depleting gas from the galaxy at 1.4$\times$ the star formation rate. These phases carry as much energy as is predicted to be fully radiated by the shocked gas \citep{1995ApJ...455..468D}. They need significantly more momentum than is initially contained in the hot ejecta from the recent starburst, with a required boost of $\sim$7. We suggest radiation pressure from an Eddington-limited starburst is a likely culprit \citep{2012ApJ...755L..26D,2015MNRAS.449..147T}, on top of the momentum produced by the hot wind \citep{2016MNRAS.455.1830T,2021MNRAS.504.3412L,2022ApJ...924...82F}.

The slower, Episode~I wind has a much lower outflow rate of $dM/dt_\mathrm{I}^\mathrm{HII} \sim 10~\mathrm{M}_\odot$~yr$^{-1}$. However, this ionized wind has moved into the galaxy's CGM and is depositing gas in a much larger volume. It contains even more ionized gas; in our density model, $M_\mathrm{I}^\mathrm{HII} = 5\times10^9$~M$_\odot$. This increase in mass as the wind slows but moves to larger radius may be due to loading from radiative cooling of the hot wind or CGM and/or mixing of the hot gas into the cool phase through mixing layers \citep{2018MNRAS.481.1873L,2020MNRAS.492.1970G,2020ApJ...894L..24F}. The gas may also be shocking on the surrounding CGM \citep{2016MNRAS.455.1830T}. The energy contained in this flow is quite small, however, compared to the flux predicted to be radiated in the shocked gas, with $dE/dt_\mathrm{I} = 0.003L_\mathrm{rad}$. Thus, much of the energy and momentum in Episode I may be carried in a hotter phase. Alternatively, significant energy from the fastest-moving gas in the earlier wind may simply have escaped to large radius.

The clearest way to make progress on the Makani wind and its connection to the CGM will be to try to detect hot gas in the extended nebula and measure its mass and energy content. This could be in the form of deep X-ray observations or UV observations of ionization states like \ion{O}{6}. On the other side of the energetic spectrum, the dusty, neutral phase is detected to 20--25~kpc, but its physical conditions are mostly unknown. Deep mid-infrared imaging and spectroscopy with the {\it JWST} would produce much stronger constraints on its presence and properties in both wind episodes. Finally, we could reduce uncertainties in the present work through integral-field maps of the recombination lines in Makani with wide-field, red-sensitive instruments like the Multi Unit Spectroscopic Explorer (MUSE) or the Keck Cosmic Reionization Mapper (KCRM).

\begin{acknowledgements}

We thank Drummond Fielding and Cassi Lochhaas for insights into recent wind models, and Sylvain Veilleux for comments on the manuscript. We acknowledge support from the National Science Foundation (NSF) under a collaborative grant (AST-1813299, 1813365, 1814233, 1813702, and 1814159). S.P. and A.L.C. acknowledge support from the Heising-Simons Foundation grant 2019-1659.

The data presented herein were obtained at the W. M. Keck Observatory, which is operated as a scientific partnership among the California Institute of Technology, the University of California and the National Aeronautics and Space Administration. The Observatory was made possible by the generous financial support of the W. M. Keck Foundation. 

The authors wish to recognize and acknowledge the very significant cultural role and reverence that the summit of Maunakea has always had within the indigenous Hawaiian community.  We are most fortunate to have the opportunity to conduct observations from this mountain. 

\end{acknowledgements}

\facility{Keck:II (ESI)} 
\software{ESIRedux (\url{https://www2.keck.hawaii.edu/inst/esi/ESIRedux/index.html}); IFSFIT \citep{2014ascl.soft09005R}}

\bibliography{makani_esi}{}
\bibliographystyle{aasjournal}

\setcounter{figure}{1}
\begin{figure}
    \centering
    \includegraphics[width=\textwidth]{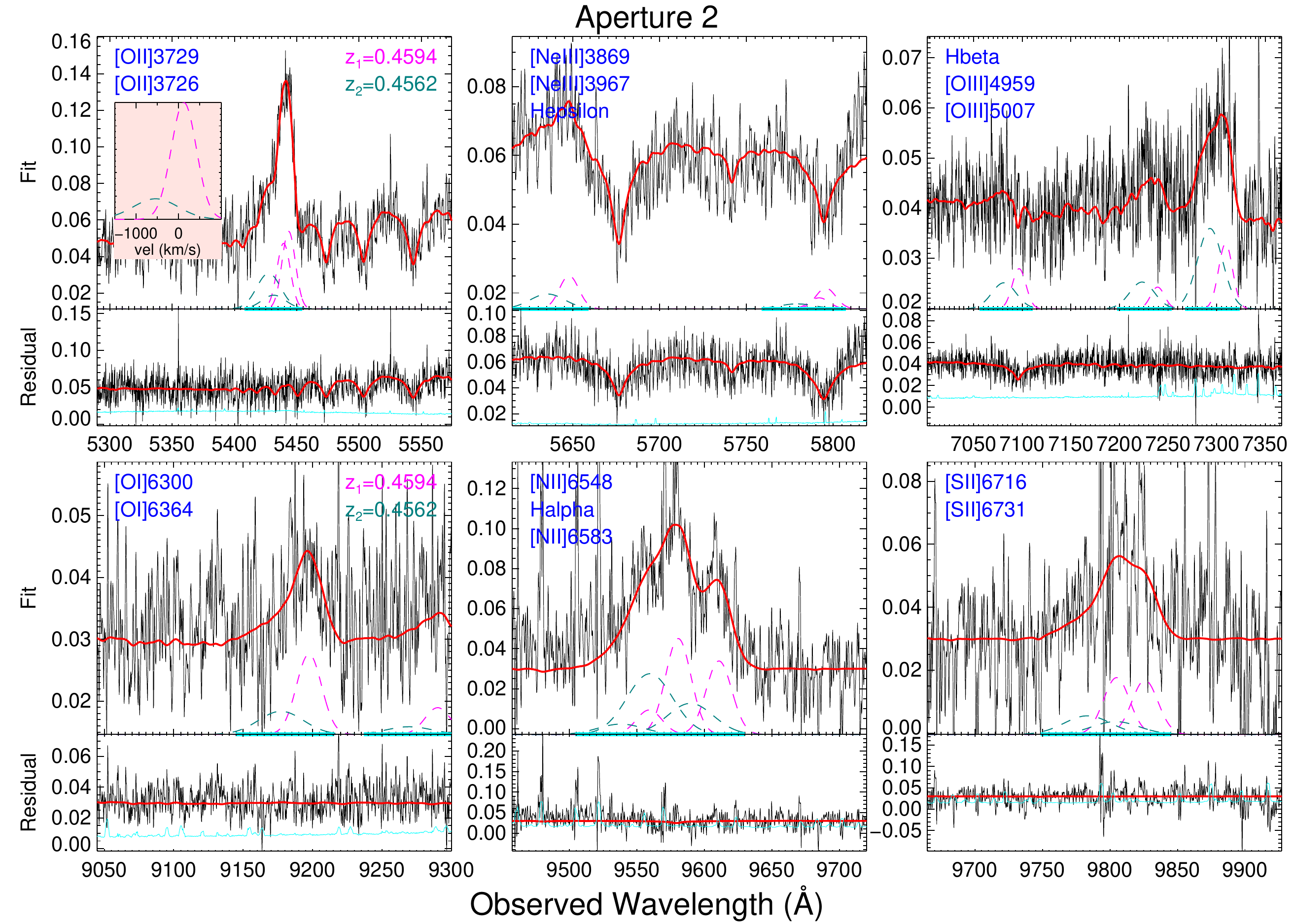}
    \caption{\it Continued.}
\end{figure}
\setcounter{figure}{1}
\begin{figure}
    \centering
    \includegraphics[width=\textwidth]{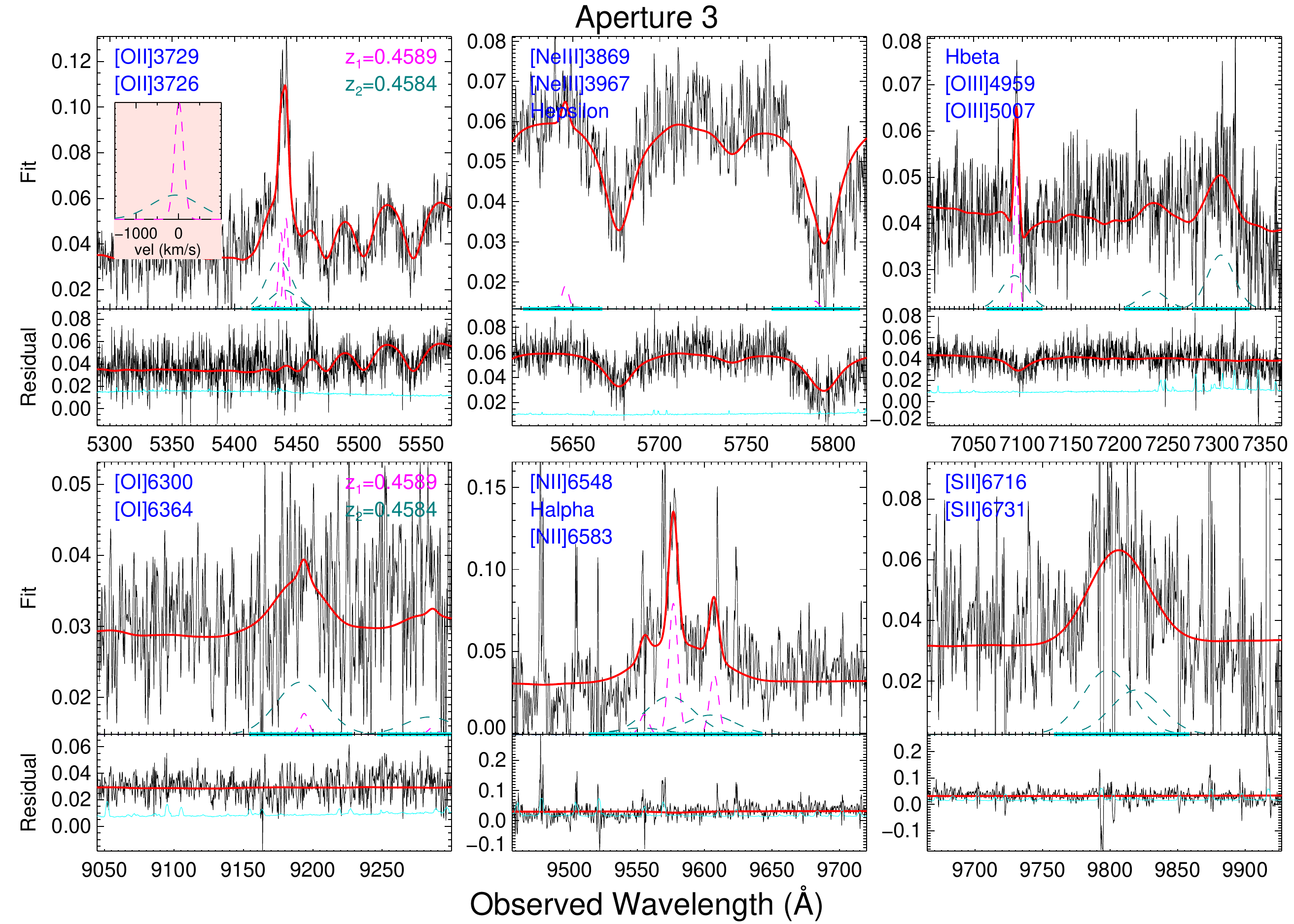}
    \caption{\it Continued.}
\end{figure}
\setcounter{figure}{1}
\begin{figure}
    \centering
    \includegraphics[width=\textwidth]{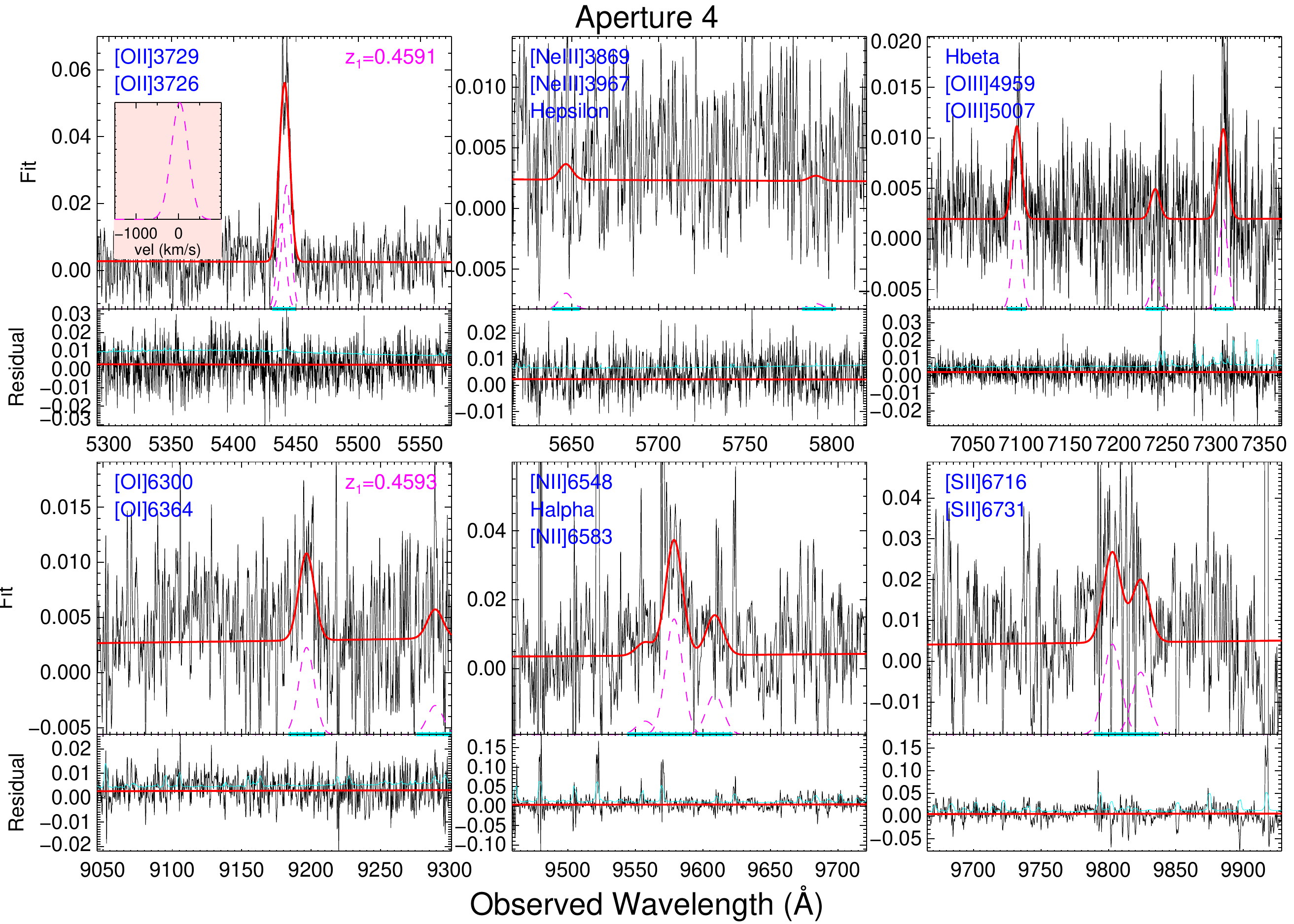}
    \caption{\it Continued.}
\end{figure}
\setcounter{figure}{1}
\begin{figure}
    \centering
    \includegraphics[width=\textwidth]{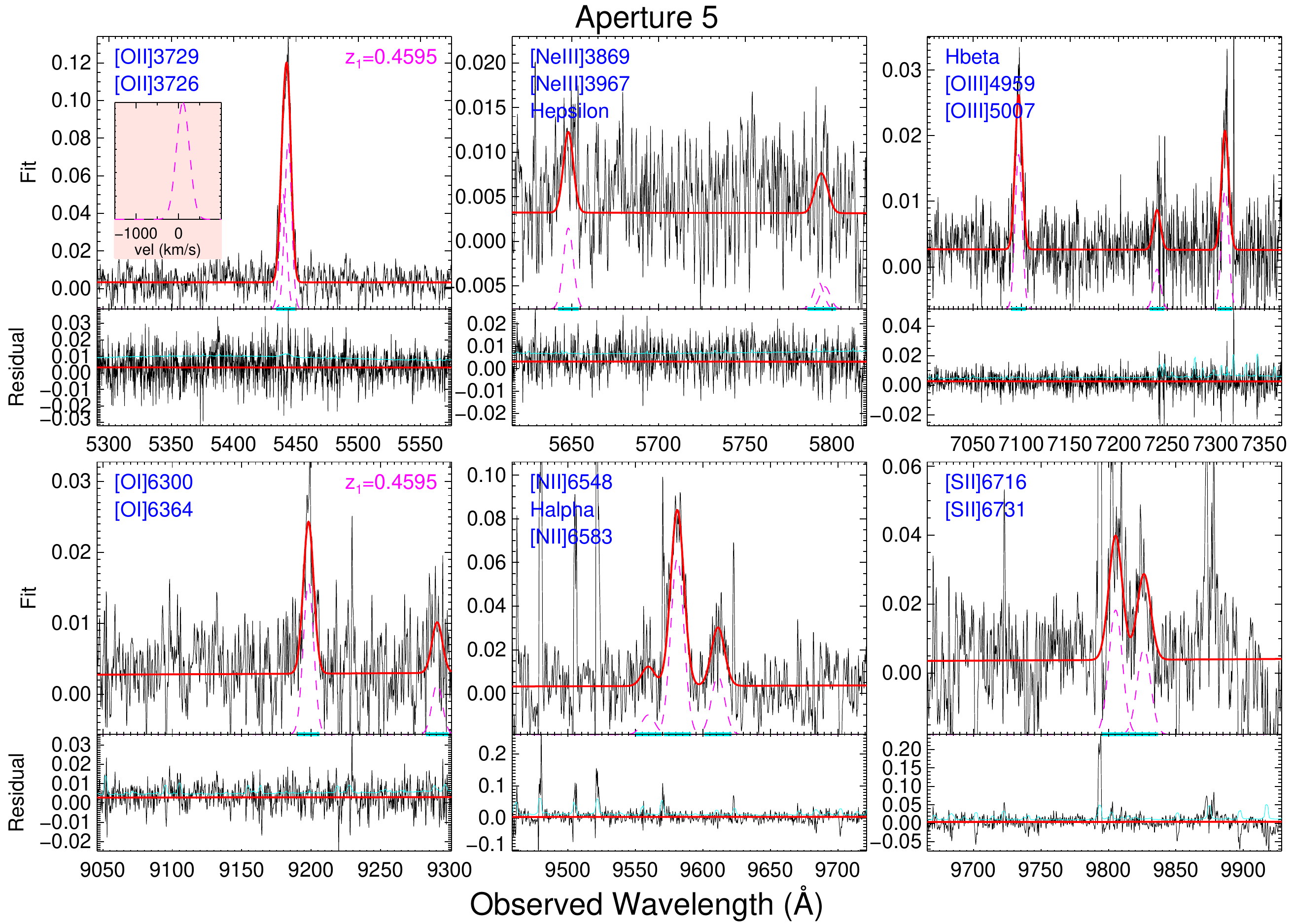}
    \caption{\it Continued.}
\end{figure}
\setcounter{figure}{1}
\begin{figure}
    \centering
    \includegraphics[width=\textwidth]{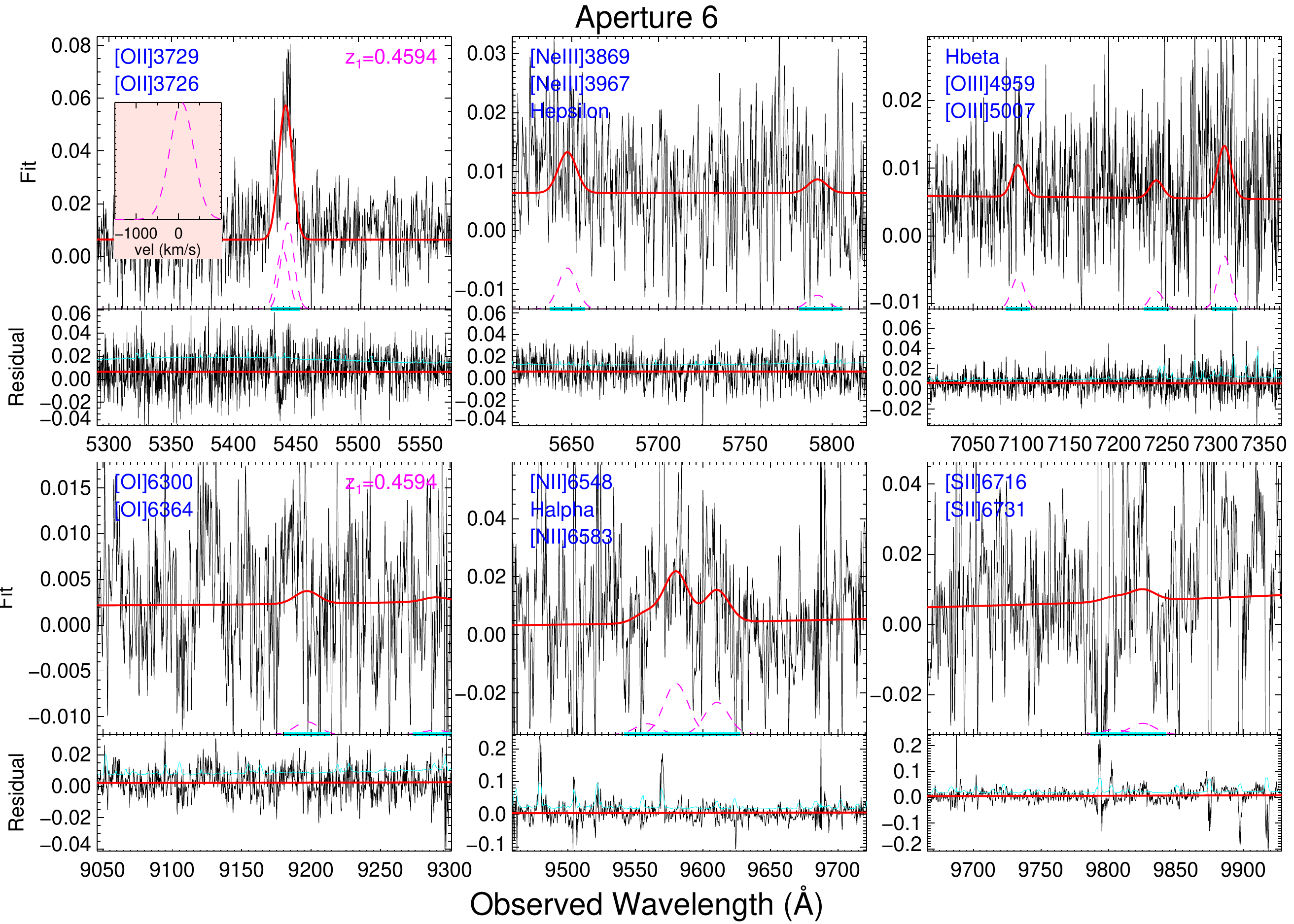}
    \caption{\it Continued.}
\end{figure}
\setcounter{figure}{1}
\begin{figure}
    \centering
    \includegraphics[width=\textwidth]{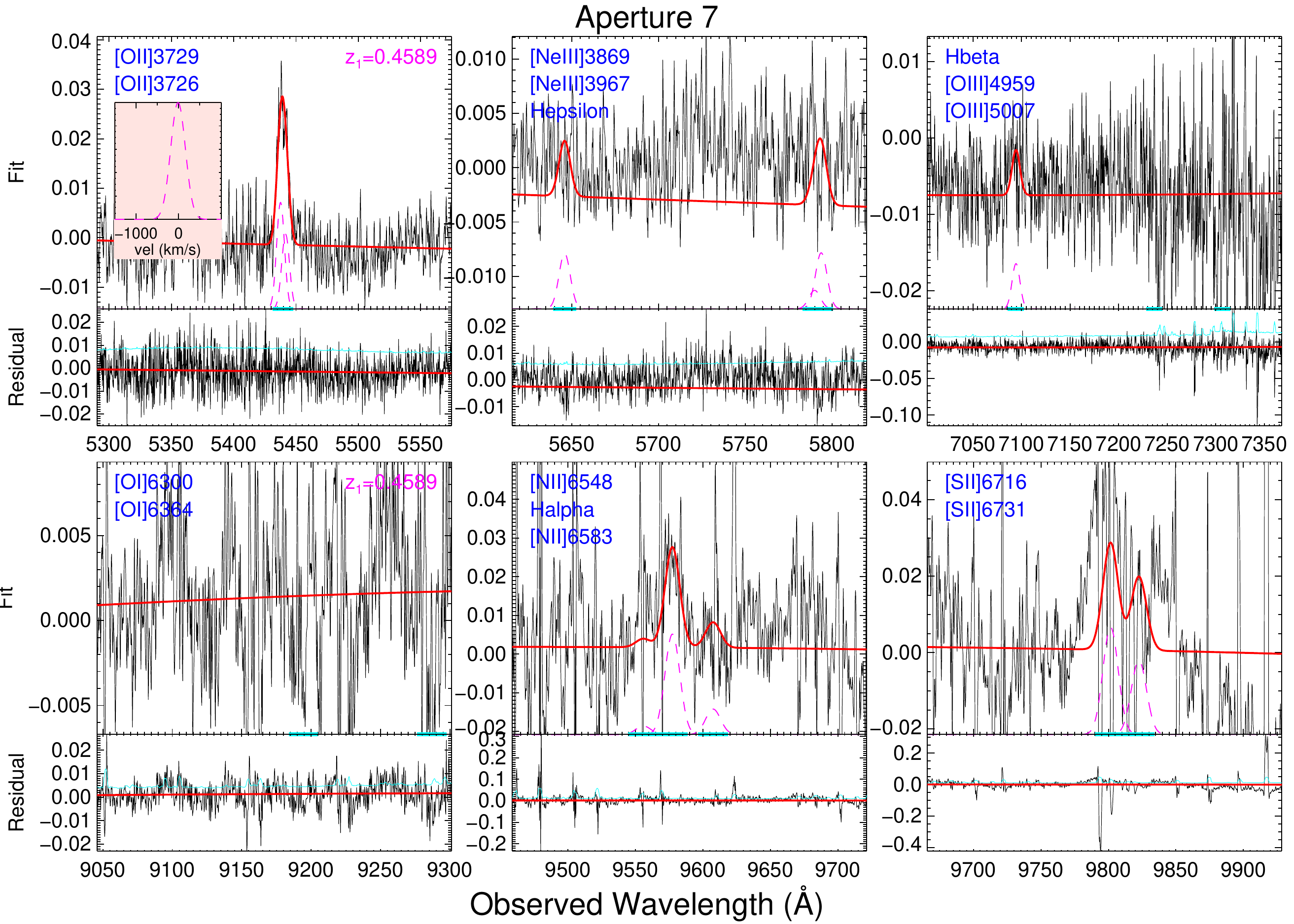}
    \caption{\it Continued.}
\end{figure}
\setcounter{figure}{1}
\begin{figure}
    \centering
    \includegraphics[width=\textwidth]{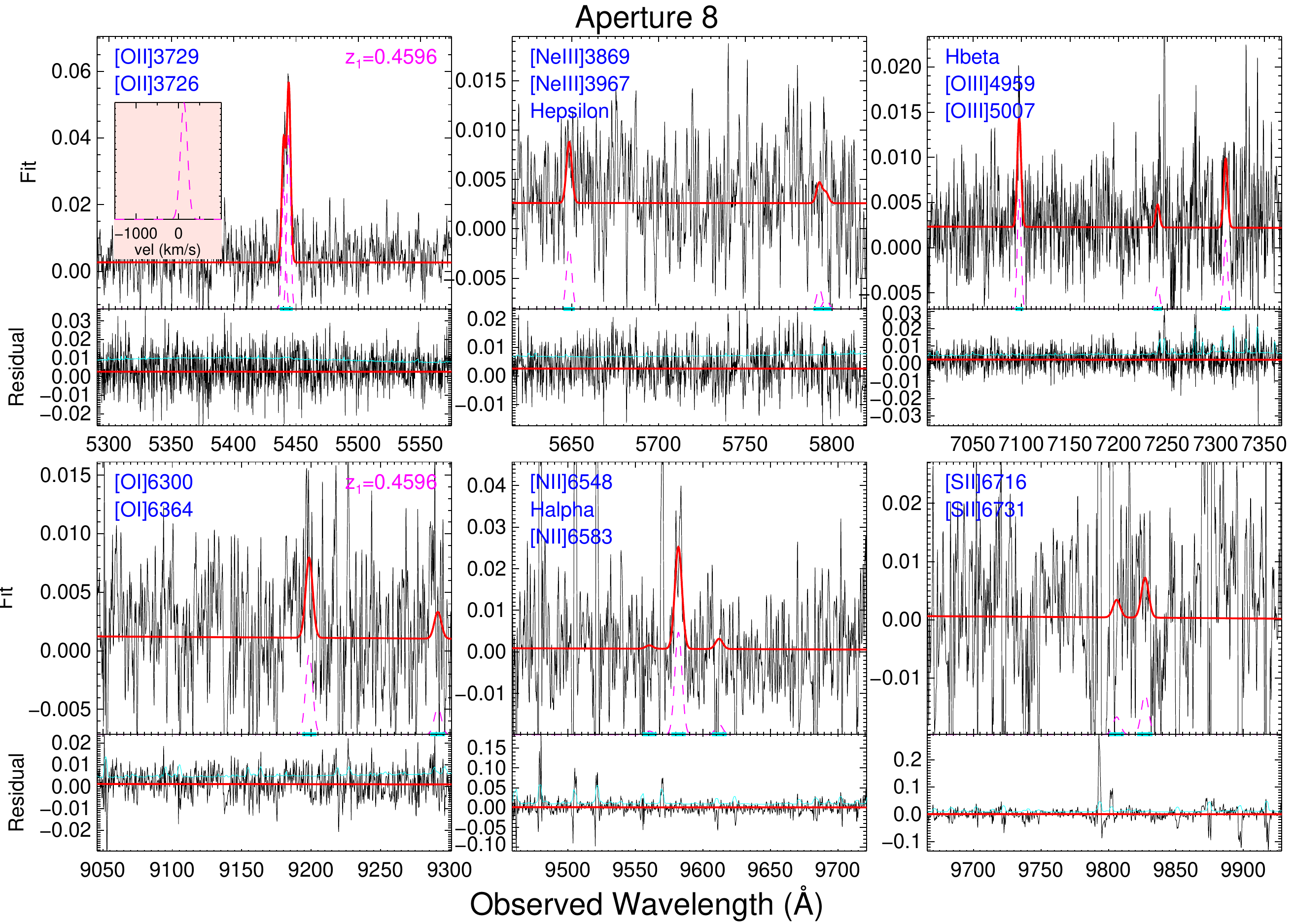}
    \caption{\it Continued.}
\end{figure}
\setcounter{figure}{1}
\begin{figure}
    \centering
    \includegraphics[width=\textwidth]{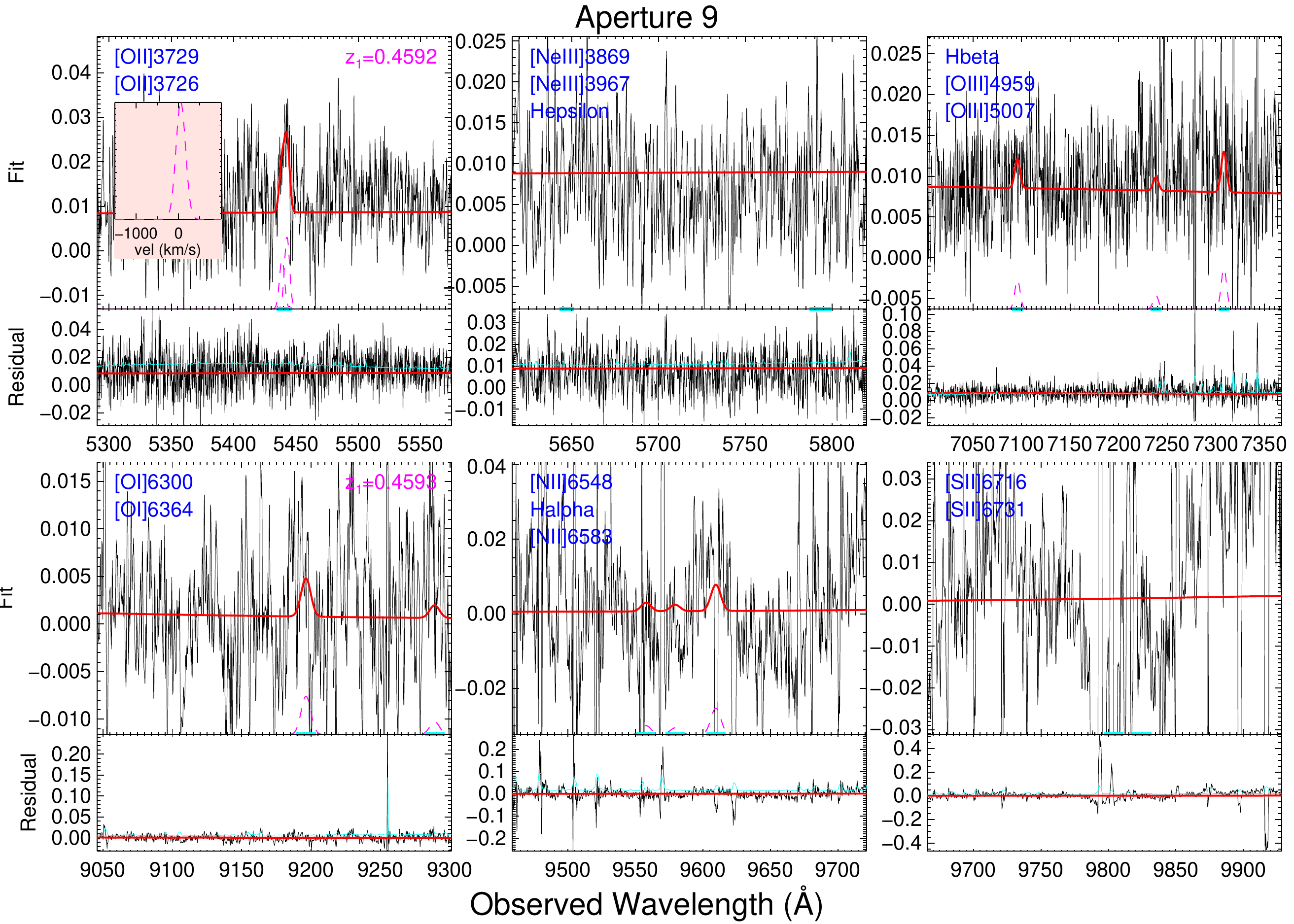}
    \caption{\it Continued.}
\end{figure}

\end{document}